\documentstyle[12pt]{article}
\def\lsim{\mathrel{\rlap {\raise.5ex\hbox{$ < $}}
{\lower.5ex\hbox{$\sim$}}}}

\newcommand{\pr}{\paragraph{}}
\newcommand{\be}{\begin{equation}}
\newcommand{\ee}{\end{equation}}
\newcommand{\bea}{\begin{eqnarray}}
\newcommand{\nn}{\nonumber}
\newcommand{\eea}{\end{eqnarray}}
\newcommand{\nd}[1]{/\hspace{-0.6em} #1}
\newcommand{\nk}{\noindent}
\def\gappeq{\mathrel{\rlap {\raise.5ex\hbox{$>$}}
{\lower.5ex\hbox{$\sim$}}}}

\def\lappeq{\mathrel{\rlap{\raise.5ex\hbox{$<$}}
{\lower.5ex\hbox{$\sim$}}}}

\begin{document}

\begin{titlepage}
\begin{flushright}
ACT-19/95 \\
CTP-TAMU-55/95 \\
OUTP-95-52P     \\
quant-ph/9512021 \\
\end{flushright}

\begin{centering}
\vspace{.1in}
{\large {\bf
A Non-Critical String (Liouville) Approach
to Brain Microtubules: State Vector Reduction,
Memory Coding and Capacity
}} \\
\vspace{.2in}
{\bf N.E. Mavromatos$^{a,\diamond}$} and
{\bf D.V. Nanopoulos$^{b}$}
\vspace{.2in}
\begin{flushleft}
$^{a}$ Dept. of Physics
(Theoretical Physics), University of Oxford, 1 Keble Road,
Oxford OX1 3NP, U.K.  \\
$^{b}$ Texas A \& M University, College Station, TX 77843-4242, USA
and Astroparticle Physics Group, Houston
Advanced Research Center (HARC), The Mitchell Campus,
Woodlands, TX 77381, USA. \\
\end{flushleft}
\vspace{.03in}

{\bf Abstract} \\
\vspace{.1in}
\end{centering}
{\small Microtubule (MT) networks, subneural
paracrystalline cytosceletal structures,
seem to play a fundamental role in the
neurons. We cast here the complicated
MT dynamics in the form of a $1+1$-dimensional
non-critical string theory, thus enabling
us to provide a consistent quantum
treatment of MTs, including enviromental
{\em friction} effects. We suggest, thus,
that the MTs are the microsites, in
the brain, for the emergence of stable, macroscopic
quantum coherent states,
identifiable with the {\em preconscious states}.
Quantum space-time effects, as
described by non-critical string theory, trigger
then an {\em organized collapse}
of the coherent states
down to a specific or {\em conscious state}. The whole
process we estimate to
take ${\cal O}(1\,{\rm sec})$, in excellent agreement
with a plethora of experimental/observational
findings. The {\em microscopic
arrow of time}, endemic
in non-critical string theory, and
apparent here in
the self-collapse process, provides
a satisfactory and simple resolution to
the age-old problem of how the, central to our feelings
of awareness, sensation
of the progression of time is generated. In addition,
the complete integrability of the stringy model for MT
we advocate in this work proves sufficient
in providing a satisfactory solution
to memory coding and capacity. Such features
might turn out to be important
for a model of the brain as a quantum computer.}

\vspace{0.1in}
\begin{flushleft}
December 1995  \\
\end{flushleft}
\vspace{.01in}
\pr
\nk $^{\diamond}$ P.P.A.R.C. Advanced Fellow

\end{titlepage}
\newpage
\section{Introduction}
\pr
The interior of living cells is structurally
and dynamically organized by {\it cytoskeletons}, i.e.
networks of protein polymers. Of these structures,
{\it MicroTubules} (MT) appear to be \cite{hameroff}
the most fundamental. Their dynamics has been
studied recently by a number of authors in connection
with the mechanism responsible for
dissipation-free
energy transfer. Recently,
Hameroff and Penrose \cite{HP} have conjectured another
fundamental r${\hat o}$le for the MT, namely being
responsible for {\it quantum computations}
in the human brain, and, thus, related to the
consciousness of the human mind.  The latter is argued to be
associated with certain aspects of quantum theory \cite{penrose,dn}
that are believed to occur in the cytoskeleton MT, in particular
quantum superposition and subsequent collapse of the
wave function of coherent MT networks.
While quantum superposition is a well-established and well-understood
property of quantum physics, the collapse of the  wave function has been
always enigmatic. We propose here to use an explicit string-derived
mechanism -
in one {\it interpretation} of non-critical string theory -
for the collapse of the  wave function\cite{emn},
involving quantum
gravity in an
essential way and solidifying previous intuitively plausible
suggestions\cite{ehns,emohn}.
It is an amazing surprise that quantum gravity effects, of
order of magnitude $G_N^{1/2}m_{p}\sim10^{-19}$, with $G_N$ Newton's
gravitational constant and $m_{p}$ the proton mass,
can play a r\^ole in such low energies as the $eV$ scales
of the typical energy transfer that occurs in cytoskeleta.
However,
as we show in this article, the fine details of the MT characteristic
 structure
indicate that not only is this conceivable, but such
scenaria
lead to order of magnitude estimates for the
time scales entering {\it conscious perception}  that are close enough
to those conjectured/``observed'' by neuroscientists,
based on completely different grounds.
\pr
To understand how quantum space-time effects can affect
conscious perception, we mention that
it has long been suspected \cite{Frohlich}
that large scale quantum coherent phenomena can occur
in the interior of biological cells, as a result of the existence
of ordered water molecules. Quantum mechanical vibrations
of the latter are
responsible for the appearance of
`phonons' similar in nature to those associated with
superconductivity. In fact there is a close analogy between
superconductivity and energy transfer in biological cells.
In the former
phenomenon electric current is transferred without dissipation
in the surface of the superconductor. In biological cells, as we shall
discuss later on, energy is transferred through the cell
without {\it loss},
despite the existence of frictional forces that represent
the interaction of the cell with the surrounding water
molecules \cite{lal}.
Such large scale quantum coherent states can maintain
themselves for up to ${\cal O}(1\,{\rm sec})$,
without significant
environmental entanglement. After that time, the state
undergoes self-collapse, probably
due to quantum
gravity effects. Due to quantum transitions
between the different states
of the quantum system
of MT in certain parts of the human brain,
a sufficient distortion of the surrounding space-time
occurs,
so that a microscopic (Planck size) black hole is formed.
Then collapse is induced, with a collapse time that
depends on the order of magnitude of the number $N$ of
coherent
microtubulins. It is estimated that, with an $N=O[10^{12}]$,
the collapse time
of
${\cal O}(1\,{\rm sec})$,
which appears to be
a typical time scale of conscious events, is achieved.
Taking into account that experiments have shown that
there exist
$N=10^{8}$ tubulins per neuron, and that there are $10^{11}$
neurons in the brain,
it follows that
that this order of magnitude for $N$ refers to a
fraction $10^{-7}$ of the human brain, which is very close to
the fraction believed responsible for human perception.
\pr
The self-collapse of the MT coherent state  wave function
is an essential step for the operation of the MT network
as a quantum computer. In the past it has been suggested
that MT networks processed information in a way similar
to classical cellular automata (CCA)
\cite{hamcel}. These
are described by interacting Ising spin chains on the spatial
plane obtained by fileting open and flattening the MT cylindrical
surface. Distortions in the configurations of individual
parts of the spin chain can be influenced by the environmental spins,
leading to information processing.
In view of the suggestion
\cite{HP} on viewing the conscious parts of the human mind
as quantum computers,
one might extend the concept of the
CCA to a quantum cellular automaton (QCA), undergoing
wave function self-collapses due to (quantum gravity)
enviromental entanglement.
\pr
An interesting and basic
issue that
arises in connection with the above r\^ole of the
brain as a quantum computer
is the emergence of a
direction in the flow of time (arrow). The
latter could be the result
of succesive self-collapses of the system's  wave function.
In a recent series of papers
\cite{emn} we have suggested a rather detailed
mechanism by which an {\it irreversible}
time variable has emerged in certain models of string  quantum
gravity. The model utilized string particles propagating
in singular space-time backgrounds with event horizons.
Consistency of the string approach requires conformal invariance
of the associated $\sigma$-model, which in turn implies a
coupling of the backgrounds for the propagating string modes
to an infinity of
global (quasi-topological) delocalized modes
at higher (massive)
string levels.
The existence of such couplings is necessitated by
specific coherence-preserving target space gauge symmetries
that mix the string levels \cite{emn}.
\pr
The specific model of ref. \cite{emn}
is a completely
integrable string theory, in the sense of being
characterised by an infinity of conserved charges.
This can be intuitively understood by the fact that
the model is a $(1+1)$-dimensional
Liouville string, and as such it can be mapped
to a theory of essentially free fermions
on a discretized world sheet
(matrix model approach \cite{matrix}).
A system of free fermions in $(1+1)$ dimensions
is trivially completely integrable, the infinity of
the conserved charges being provided by
appropriate moments of the fermion energies above the
Fermi surface. Of course, formally, the
precise symmetries of the model used in
ref. \cite{emn}
are much more
complicated \cite{bakas}, but the idea behind the
model's integrability is essentially the above.
It is our belief that
this quantum integrability is a very
important feature of theories of space-time
associated with the time arrow.
In its presence, theories with singular backgrounds
appear consistent as far as maintainance of
quantum coherence is concerned.
This is due to the fact that the phase-space density of the
field theory associated with the matter
degrees of freedom evolves with time according to the
conventional Liouville theorem\cite{emn}
\be
   \partial _t \rho = -\{\rho , H\}_{PB}
\label{one}
\ee
as a consequence of phase-space
volume-preserving symmetries.
In the two-dimensional example of ref. \cite{emn},
these symmetries are known as $W_{\infty}$, and are
associated
with higher spin target-space states\cite{bakas}. They are
responsible for string-level mixing, and hence they are
broken in any low-energy approximation.
If the concept of `measurement
by local scattering experiments' is introduced \cite{emn},
it becomes clear
that the observable
background cannot contain such global modes. The latter have to
be integrated out
in any effective
low-energy theory. The result of this integration
is a non-critical string theory,
based on the propagating modes only.
Its conformal invariance on the world sheet is  restored
by dressing the matter backgrounds by the
Liouville mode $\phi$, which plays the role of the time coordinate.
 The $\phi$ mode is a dynamical local
world-sheet scale \cite{emn},
flowing irreversibly as a result of certain theorems
of the renormalization group of unitary $\sigma$-models
\cite{zam}.
In this way time in target space
has a natural arrow for very specific {\it stringy reasons}.
\pr
Given the suggestion of ref. \cite{penrose}
that space-time environmental entanglement could be
responsible for conscious brain function, it is natural to
examine the conditions under which our theory \cite{emn}
can be applied.
Our approach utilizes extra degrees of freedom, the $W_\infty$
global string modes, which are not directly
accessible to local scattering `experiments' that make
use of {\it propagating} modes only.
Such degrees of freedom carry information, in a similar spirit
to the information loss suggested by Hawking\cite{hawk}
for the quantum-black-hole case. For us, such degrees of freedom
are not exotic, as suggested in ref. \cite{Page},
but
appear {\it already} in the
non-critical String Universe \cite{emn,dn},
and as such they are considered as `purely stringy'.
In this respect, we believe that
the suggested model of consciousness, based on the
non-critical-string
formalism of ref. \cite{emn}, is physically more {\it concrete}.
The idea of using string theory instead of
point-like quantum gravity is primarily associated with the
fact that a {\it consistent} quantization of gravity
is at present possible {\it only} within the
framework of string theory. However,
there are additional reasons that make advantageous
a string formalism. These include the possibility of
construction of
a completely integrable model for $MTs$, and
the Hamiltonian
representation of dynamical problems with friction involved in
the physics of MTs. This
leads to
the possibility of a consistent ({\it mean field}) {\it
quantization}
of certain soliton solutions associated with the energy
transfer mechanism in biologcal cells.
\pr
According to our previous discussion
emphasizing the importance of strings,
it is imperative that
we try to
identify the completely integrable
system underlying MT networks. Thus,
it appears essential to review
first the classical model for energy transfer
in biological systems associated with MT. This will allow
the identification of the
analogue of the (stringy)
propagating degrees of freedom, which
eventually couple to quantum (stringy) gravity
and to global environmental modes.
As we shall argue in subsequent sections, the relevant basic
building blocks of the human brain are one-dimensional
Ising spin chains,
interacting among themselves in a way so as to
create a large scale quantum coherent state,
believed to be responsible
for preconscious behaviour in the model of \cite{HP}.
The system can be described in a world-sheet
conformal invariant way and is unitary.
Coupling to gravity generates deviations from conformal
invariance which lead to time-dependence, by identifying
time with the Liouville field on the world sheet.
The situation
is similar to the environemtnal entanglement
of ref. \cite{vernon,cald}.
Due to this entanglement,
the system of the propagating modes opens up as in
Markov processes \cite{davidoff}.
This
leads to a dynamical self-collapse of the wave function
of the MT quantum coherent network.
In this way, the part of the human brain associated with
conscioussness generates, through successive collapses,
an arrow of time.
The
interaction among the spin chains,
then, provides a
mechanism for quantum computation,
resembling a planar
cellular automaton.
Such operations sustain the
irreversible flow of time.
\pr
The structure of the article is as follows:
in section 2 we discuss a model used for the
physical description of a MT, and in particular
for a
simulation of the energy transfer mechanism.
The model
can be expressed in terms of a $1+1$-dimensional
{\it classical field}, the projection of the
displacement field of
the MT dimers along the tubulin axis.
There exists {\it friction} due to intreraction
with the environment. However,
the theory possesses travelling-wave solitonic
states
responsible for loss-free transfer of energy.
In section 3, we give a formal representation
of the above system as a $1+1$-dimensional $c=1$
Liouville (string) theory. The advantage
of the method lies in that it allows for a
canonical quantization of the friction problem,
thereby yielding a model for a large-scale coherent state,
argued to simulate the preconscious state.
There is no time arrow in the above system.
In section 4, we discuss our mechanism
of introducing a dynamical time variable
with an arrow into the system, by elevating the
above $c=1$ Liouville theory to a $c=26$ non-critical
string theory, incorporating quantum
gravity effects. Such effects arise from the distortion
of space-time due to abrupt conformational
changes in the dimers. Such a coupling
leads
to a breakdown of the quantum coherence
of the preconscious state.
Estimates of the collapse times are given,
with the result that in this approach
concious perception of a time scale of
${\cal O}(1\,{\rm sec})$,
is due to a $10^{-7}$ part of the total brain.
In section 5 we briefly discuss
{\it growth} of a MT network in our framework,
which would be the analogue of
a non-critical string driven inflation
for the effective one-dimensional
universe of the MT dimer degrees of freedom.
We view MT {\it growth }
as an out-of-equilibrium one-dimensional
spontaneous-symmetry breaking process
and discuss the connection of
our approach to
some elementary theoretical
models with driven diffusion
that could serve as prototypes for
the phenomenon. We also point out some
other experiments, sensitive to weak fields -
like gravitational ones, where our ideas about
a quantum integrable model for MT may find
some applications. Section 6 deals with
some interesting consequences of the complete integrability
of the model, as far as memory coding and
capacity are concerned. Both problems are resolved,
as a result of huge `stringy' symmetries,
which appear naturally and are essential in maintaining
quantum coherence of the full system, including back-reaction
effects of matter on the effective two-dimensional space-time.
Conclusions and outlook are presented in section 7.
We discuss some technical apsects of our approach
in two
Appendices.

\section{Physical Description of the Microtubules}
\subsection{The model and its parameters}
\pr
In this section
we review certain features
of the MT that will be useful in subsequent parts
of this work. MT are hollow cylinders (cf Fig 1)
comprised of an exterior surface
(of cross-section
diameter
$25~nm$)
with 13 arrays
(protofilaments)
of protein
dimers
called tubulins.
The interior of the cylinder
(of cross-section
diameter $14~nm$)
contains ordered water molecules,
which implies the existence
of an electric dipole moment and an electric field.
The arrangement of the dimers is such that, if one ignores
their size,
they resemble
triangular lattices on the MT surface. Each dimer
consists of two hydrophobic protein pockets, and
has an unpaired electron.
There are two possible positions
of the electron, called $\alpha$ and $\beta$ {\it conformations},
which are depicted in Fig. 2. When the electron is
in the $\beta$-conformation there is a $29^o$ distortion
of the electric dipole moment as compared to the $\alpha $ conformation.
\pr
In standard models for the simulation of the MT dynamics,
the `physical' degree of freedom -
relevant for the description of the energy transfer -
is the projection of the electric dipole moment on the
longitudinal symmetry axis (x-axis) of the MT cylinder.
The $29^o$ distortion of the $\beta$-conformation
leads to a displacement $u_n$ along the $x$-axis,
which is thus the relevant physical degree of freedom.
This way, the effective system is one-dimensional (spatial),
and one has a first indication that quantum integrability
might appear. We shall argue  later on
that this is indeed the case.
\pr
Information processing
occurs via interactions among the MT protofilament chains.
The system may be considered as similar to a model of
interacting Ising chains on a trinagular lattice, the latter being
defined on the plane stemming from fileting open and flatening
the cylindrical surface of Fig. 1.
Classically, the various dimers can occur in either $\alpha$
or $\beta$ conformations. Each dimer is influenced by the neighboring
dimers resulting in the possibility of a transition. This is
the basis for classical information processing, which constitutes
the picture of a (classical) cellular automatum.
\pr
The quantum computer character of the MT network results
from the assumption that each dimer finds itself in a
superposition of $\alpha$ and $\beta$ conformations \cite{HP}.
There is a macroscopic
coherent state among the various chains, which
lasts for ${\cal O}(1\,{\rm sec})$ and constitutes the `preconscious'
state. The interaction
of the chains with (stringy) quantum gravity, then, induces
self-collapse of the wave function of the coherent MT network,
resulting in quantum computation.
\pr
In what follows we shall assume that the collapse occurs
mainly due to the interaction of each chain with
quantum gravity, the interaction from neighboring chains
being taken into account by including mean-field interaction terms
in the dynamics of the displacement field of each chain. This amounts
to a modification of the effective potential by
anharmonic oscillator terms.
Thus, the effective system under study is two-dimensional,
possesing one space and one time coordinate. The precise
meaning of `time' in our model will be clarified when
we discuss the `non-critrical string' representation
of our system.
\pr
Let $u_n$ be the displacement field of the $n$-th dimer in a MT
chain.
The continuous approximation proves sufficient for the study of
phenomena associated with energy transfer in biological cells,
and this implies that one can make the replacement
\be
  u_n \rightarrow u(x,t)
\label{three}
\ee
with $x$ a spatial coordinate along the longitudinal
symmetry axis of the MT. There is a time variable $t$
due to
fluctuations of the displacements $u(x)$ as a result of the
dipole oscillations in the dimers.
At this stage, $t$ is viewed as a reversible variable.
The effects of the neighboring
dimers (including neighboring chains)
can be phenomenologically accounted for by an effective
double-well potential \cite{mtmodel}
\be
U(u) = -\frac{1}{2}A u^2(x,t) + \frac{1}{4}Bu^4(x,t)
\label{four}
\ee
with $B > 0$. The parameter $A$ is temperature
dependent. The
model of ferroelectric distortive spin chains
of ref. \cite{collins} can be used to
give a temperature
dependence
\be
   A =-|const|(T-T_c)
\label{temper}
\ee
where $T_c$ is a critical temperature of the system, and
the constant is determined phenomenologically \cite{mtmodel}.
In realistic cases the temperature $T$ is very close to $T_c$,
which for the human brain is taken to be the room temperature
$T_c = 300K$.
Thus, below $T_c$
$A > 0$.
The important relative minus sign in the potential (\ref{four}), then,
guarantees the necessary degeneracy, which is necessary for the
existence of
classical solitonic solutions. These constitute the basis
for our coherent-state description of the
preconscious state.
\pr
Including a phenomenological kinetic term for the dimers,
each having a mass $M$, one can write down a Hamiltonian
\cite{mtmodel}
\be
H = kR_0^2 (\partial _x u)^2  - M (\partial _t u)^2
-\frac{1}{2}A u^2 + \frac{1}{4}B u^4
+ qEu
\label{five}
\ee
where $k$ is a stiffness parameter, $R_0$ is the equilibrium
spacing between adjacent dimers, $E$ is the electric field
due to the `giant dipole' representation of the MT cylinder,
as suggested by the experimental results \cite{mtmodel},
and $q=18 \times 2e$ ($e$ the electron charge) is a
mobile charge. The spatial-derivative term in (\ref{five})
is a continuous approximation of terms in the lattice Hamiltonian
that express the effects of restoring strain forces between
adjacent dimers in the chains \cite{mtmodel}.
\pr
The effects of the surrounding water molecules can be
summarized by a viscuous force term that damps out the
dimer oscillations,
\be
 F=-\gamma \partial _t u
\label{six}
\ee
with $\gamma$ determined phenomenologically at this stage.
This friction should be viewed as an environmental effect, which
however does not lead to energy dissipation, as a result of the
non-trivial
solitonic structure of the
ground-state
and the non-zero constant
force due to the electric field.
This is a well known result, directly relevant to
energy transfer in biological systems \cite{lal}.
The modified equations of motion, then, read
\be
    M \frac{\partial ^2 u }{\partial t^2}
- k R_0^2 \frac{\partial ^2  u}{\partial x^2} - A u
+ B u^3 + \gamma \frac{\partial u}{\partial t} - qE =0
\label{seven}
\ee
According to ref. \cite{lal} the importance
of the force term $ qE $ lies in the fact
that eq (\ref{seven}) admits displaced classical soliton solutions
with no energy loss.
The solution acquires the form of a travelling wave,
and can
be most easily exhibited by defining a normalized
displacement field
\be
  \psi (\xi ) = \frac{u (\xi)}{\sqrt{A/B}}
\label{eight}
\ee
where,
\be
               \xi \equiv \alpha( x - v t)
\qquad \alpha \equiv \sqrt{\frac{|A|}{M(v_0^2 - v^2)}}
\label{nine}
\ee
with
\be
v_0 \equiv \sqrt{k/M} R_0
\label{sound}
\ee
the
sound velocity, of order $1 km/sec$,
and $v$ the propagation velocity
to be determined later.
In terms of the $\psi (\xi )$ variable,
equation (\ref{seven}) acquires the form of the equation
of motion of an anharmonic oscillator in a frictional environment
\bea
 \psi '' &+& \rho \psi ' - \psi ^3 + \psi + \sigma = 0  \nn \\
\rho &\equiv& \gamma v [M |A|(v_0^2 - v^2)]^{-\frac{1}{2}}, \qquad
\sigma = q \sqrt{B}|A|^{-3/2}E
\label{ten}
\eea
which has a {\it unique} bounded solution \cite{mtmodel}
\be
    \psi (\xi ) = a + \frac{b -a}{1 + e^{\frac{b-a}{\sqrt{2}}\xi}}
\label{eleven}
\ee
with the parameters $b,a$ and $d$ satisfying:
\be
(\psi -a )(\psi -b )(\psi -d)=\psi ^3 - \psi -
\left(\frac{q \sqrt{B} }{|A|^{3/2}} E\right)
\label{twelve}
\ee
Thus, the kink propagates along the protofilament axis
with fixed velocity
\be
    v=v_0 [1 + \frac{2\gamma^2}{9d^2M|A|}]^{-\frac{1}{2}}
\label{13}
\ee
This velocity depends on the strength of the electric
field $E$ through the dependence of $d$ on $E$ via (\ref{twelve}).
Notice that, due to friction, $v \ne v_0$, and this is essential
for a non-trivial second derivative term in (\ref{ten}), necessary
for wave propagation.
For realistic biological systems $v \simeq 2 m/sec$.
With a velocity of this order,
the travelling waves
of kink-like excitations of the displacment field
$u(\xi )$ transfer energy
along a moderately long microtubule
of length $L =10^{-6} m$ in about
\be
t_T = 5 \times 10^{-7} sec
\label{transfer}
\ee
This time is very close to Frohlich's time for
coherent phonons in biological system.
We shall come back to this issue later on.
\pr
The total energy of the solution (\ref{eleven}) is
easily calculated to be \cite{mtmodel}
\be
   E = \frac{1}{R_0} \int _{-\infty}^{+\infty} dx H
= \frac{2\sqrt{2}}{3}\frac{A^2}{B} +  \frac{\sqrt{2}}{3}
k \frac{A}{B} + \frac{1}{2} M^{*} v^2  \equiv \Delta + \frac{1}{2}
M^{*} v^2
\label{energy}
\ee
which is {\em conserved} in time.
The `effective' mass $M^{*}$ of the kink is given
by
\be
            M^{*} = \frac{4}{3\sqrt{2}}\frac{MA\alpha }{R_0 B}
\label{effmass}
\ee
The first term of equation (\ref{energy})
expresses the binding energy of the kink
and the second the resonant transfer energy.
In realistic
biological models the sum of these two terms
dominate over the third
term, being of order of $1eV$ \cite{mtmodel}. On the other hand,
the effective mass in (\ref{effmass}) is\cite{mtmodel} of order
$5 \times 10^{-27} kg$, which is about
the proton mass ($1 GeV$) (!).
As
we shall discuss later on, these values are essential
in yielding the correct estimates for the
time of collapse of the `{\it preconscious}' state due to our
quantum gravity environmental entangling. To make plausible
a consistent
study of such effects, we now discuss the possibility of
representing the equations of motions (\ref{ten})
as being derived from string theory.
\pr
Before closing we mention that the above {\it classical}
kink-like excitations (\ref{eleven})
have been discussed so far in connection
with physical mechanisms associated with
the
hydrolysis of GTP (Guanosine-ThreePhosphate)
tubulin dimers to GDP (Guanosine-DiPhosphate) ones.
Because the two forms of tubulins correspond
to different conformations $\alpha$ and $\beta$
above, it is conceivable to speculate that
the quantum mechanical oscillations between
these two forms of tubulin dimers
might be associated with a quantum version
of kink-like excitations in the MT network.
This is the idea we put forward in the present
work. The novelty of our approach
is the use of Liouville (non-critical) string theory
for the study of the dynamics involved.
This is discussed in the next section.
\pr
Before doing so we consider it as useful to
discuss the r\^ole of the surrounding water molecules
as a medium providing us with the initial quantum
coherent `preconscious' state of mind.
We shall briefly review existing works on the subject
and outline the advantages of our approach, which uses
completely integrable (non-critical string)
theories to represent the MT dynamics.

\subsection{The importance of the water environment}
\pr
We start by first reviewing
existing
mechanisms conjecturing
the importance of the surrounding water molecules
for the proper functioning or even the existence
of MT~\cite{delgiud}. As a result of
its electric dipole structure,
the ordered water environment exhibits a laser-like
behaviour~\cite{prep}. Coherent modes emerge as a result of the
interaction of the
electric dipole moments
of the water molecules
with the quantized electromagnetic radiation.
Such quanta can be understood~\cite{delgiud}
as Goldstone modes arising from the spontaneous
breaking of the electric dipole symmetry, which in the
work of ref. \cite{delgiud} was the only symmetry
to be assumed spontaneously broken.
In our string model,
as we shall discuss
later on, a more complicated (infinite-dimensional) symmetry
breaks spontaneously, which incorporates the simple rotational
symemtry of the point-like theory models.
The emergence of coherent dipole quanta resembles
the picture of Fr\"ohlich
coherent `phonons'~\cite{Frohlich},
emerging in biological
systems for energy transfer without dissipation.
\pr
The existence of such coherent states in the surrounding
water results in the friction term proportional to $\rho$
in (\ref{ten}). What we have argued above is that,
because of this interaction,
a kink soliton can be formed, provided that
the MT are of sufficient length. Such solitons
can then be themselves squeezed coherent states, being responsible
for a  {\it preconscious} state of the mind. This provides an
explicit mechanism for the realization of the conjectures of \cite{HP}.
Such coherent states extend over huge networks of MT
covering approximately a fraction $10^{-7}$ of the human brain.
Quantum gravity effects, then, are responsible for the collapse
of such states in MT networks, leading to `conscious' perception.
Such effects are represented by the formation of
virtual black hole states in (1+1)-dimensional
non-critical string models. As we shall discuss in subsequent
sections,
this spoils the conformal invariance
(criticality) of the
effective $\sigma$-model based only on the displacement field
$\psi$ of MT.
To restore criticality,
it is necessary to have time-dependent parts in the
configuration of the dilaton field~\cite{emn},
{\it in  addition} to the space-like linear dilaton backgrounds
characterizing (1+1)-dimensional flat space-times.
It is important to stress that space-like dilaton
backgrounds are {\it necessary} for asymptotic
flatness of space-time~\cite{witt}. Such backgrounds are
provided by the friction ($\rho$) with the surrounding
water.
There lies the importance of the existence of effects
originating from the interaction of the MT with the
surrounding
water molecules, which co-exist with quantum gravity effects
that lead to a collapse of the preconscious state.
\pr
At this point we should stress that our suggested mechanism
for conscious perception' in the framework of non-critical string
(integrable) models has similarities with, but also important
differences from,
previous mechanisms of brain functioning, suggested
in the context of local quantum
field theory~\cite{umez,stewart,delgiud}.
In the latter case, {\it external stimuli}
are believed responsible for triggering
spontaneous breaking of the electric dipole rotational
symmetry of the water environment.
The collective Goldstone modes of such a
breaking (dipole quanta)
are spin wave quanta and the
system's phases are macroscopically characterized
by the value of the order parameter, which in this case
is the polarization ({\it coding}). If the
ground state of the system is considered as the {\it memory}
state, then the above process is just {\it memory printing}.
In this picture,
{\it memory recall} corresponds to the excitation,
under another external stimulus,
of dipole quanta of similar nature to those leading
to the printing.
The brain, then,
`consciously feels'~\cite{stewart,vitiello} the pre-existing order
in the ground state.
\pr
The usual problem with such mechanisms is associated
with {\it memory capacity}. The set of code numbers
associated with the breaking of just a limited number
of symmetries is not sufficient~\cite{stewart}
to explain the
storage of an enormous amount of information in the human brain.
We shall discuss these issues later on. At the moment we only mention
the interesting suggestion of ref. \cite{vitiello} that dissipation
can lead to an evasion of the problem of memory capacity,
as a result of the doubling of the degrees of freedom
of the dissipative system in order to achieve canonical
quantization. In this case, memory states can be shown to
be classified by infinite dimensional non-compact
symmetries
(in a thermodynamic limit), leading to the existence
of ground state {\it replicas}. In this way {\it overprinting}
is avoided. In our opinion, althought such an approach
is very appealing and certainly shares many things in
common with our stringy picture for the brain MT, however
has
the disadvantage of the
non-hermiticity of the Hamiltonian~\cite{umez,umeztfd}, thus
leading to macroscopic energy dissipation.
In order to make contact with energy-loss-free transport
in biological cells~\cite{Frohlich,lal} one must
find a model, where although dissipation effects exist
(c.f. $\rho$ in (\ref{ten})), and thus memory capacity
problems are treated in a satisfactory way,
however energy is
conserved on the average, and thus Frohlich's requirements
are met. Moreover, as we shall discuss below, it would be
desirable to have
a unified framework that will incorporate realistic
quantum-gravity entanglement, by means of an `environment'
formalism. All these requirements seem to be met in
a non-critical
string picture of MT dynamics~\cite{emn}, to which we now turn
in some detail.

\section{Non-Critical (Liouville)
String Theory Representation of a MT}
\pr
For this purpose,
it is important to notice that the relative sign (+)
between the second derivative and the linear term in $\psi $
in
equation (\ref{ten}) are such that this equation
can be considered as corresponding to the
tachyon $\beta$-function equation
of
a (1 + 1)-dimensional string theory, in a
flat space-time with a dilaton field $\Phi $
linear in the {\it space-like}
coordinate $\xi $ \cite{aben},
\be
        \Phi = - \rho \xi
\label{dilaton}
\ee
Indeed, the most general form of a `tachyon' deformation
in such a string theory, compatible with
conformal invariance is that of a travelling wave \cite{polch}
$T (x ' )$ , with
\bea
  x'=\gamma _{v_s} (x - v_s t)
  \qquad &;& \qquad t'= \gamma _{v_s}(t-v_s x) \nn \\
\gamma _{v_s} \equiv (1 &-& v_s^2)^{-1/2}
\label{st7}
\eea
where $v_s$ is the propagation velocity of the string `tachyon'
background.
As argued in ref. \cite{polch} these translational invariant
configurations
are the most general
backgrounds,
consistent
with a {\it unique}
factorisation of the string $\sigma$-model
theory on a Minkowski space-time $G_{\mu\nu} = \eta _{\mu\nu}$
\be
  S= \frac{1}{4\pi \alpha '} \int d^2 z
  \left[\partial X^{\mu} {\overline \partial } X^{\nu}
G_{\mu\nu}(X) + \Phi(X) R^{(2)} + T(X)\right]
\label{st2b}
\ee
into two conformal field theories,
for the $t'$ and $x'$ fields,
corresponding to central charges
\be
c_{t'} = 1 -24 v_s^2 \gamma _{v_s}^2  \qquad ; \qquad
c_{x'} = 1 + 24 \gamma _{v_s}^2
\label{st8}
\ee
In our case (\ref{ten}), the r\^ole of the space-like
coordinate $x'$
is played by $\xi $ (\ref{nine}).
The velocity of light in this effective string
model is replaced by the sound velocity $v_0$ (\ref{sound}),
and the velocity $v_s$
is defined in terms of
the velocity $v$
of the kink (\ref{13}) by
expressing
the friction coefficient
$\rho $ in terms of the central charge deficit
(\ref{st8}) (for definitions and relevant notation
see ref. \cite{polch} and also Appendix A)
\be
  \rho = \sqrt{\frac{1}{6}(c(\xi) - 1)} = 2 \gamma _{v_s}
\label{qdef}
\ee
The space-like `boosted' coordinate $\xi$, thus, plays the r\^ole
of space in this effective/Liouville mode string theory.
\pr
Notice that with the definition (\ref{qdef}) the local-field-theory
kink solution (\ref{eleven}),
propagating with a {\it real} velocity $v$,
is mapped to a non-critical string background which propagates
with a velocity $v_s$ that could be {\it imaginary} ($v_s^2 < 0$).
Indeed,
the condition for reality of $v_s$ can be easily found
from (\ref{ten},\ref{13},\ref{qdef}) to be
\be
      d^2 \ge 8/9
\label{reality}
\ee
It can be easily seen that
for the generic values of the parameters of the MT model,
described in the previous section~\cite{mtmodel},
(\ref{reality}) is not satisfied, which implies that, when
formulated as a
non-critical string theory, the MT system corresponds
to a $1 + 1$ dimensional non-critical string
with Wick-rotated `time' $s=it$~\cite{polch},
and, therefore, corresponds to a matter central charge
that overcomes the $c=1$ barrier
\be
25 \ge c_{s'} = 1 + 24 |v_s|^2 (1 + |v_s|^2 )^{-1} \ge 1
\label{barier}
\ee
In this regime, Liouville theory is poorly understood, but
there is the belief that this range of matter central charges
is characterized by
polymerization properties
of random surfaces. From our point of view this may be related
to certain growth properties of the MT networks, which are
briefly discussed in section 5.
\pr
An interesting question arises as to whether there are
circumstances under which (\ref{reality}) is satisfied,
in which case the matter content of the theory
is characterized by the central charge $c_{t'} \le 1 $ (\ref{st8}).
To this end, we note that
the parameter $d$ in (\ref{reality}) has a large uncertainty
since, among others, it depends on $A$.
At present,
there are no accurate experimental data for
$A$, which depends on temperature (\ref{temper}).
The parameter $d$ is also sensitive
to the order of magnitude of the electric field $E$.
The latter is non-uniform along the MT axis. There is a
sharp increase of $E$ towards the end points
of the MT protofilament axis ~\cite{mtmodel},
and for such large $E$, $d \sim E^{1/3}$.
Due to
(\ref{13}), an increase in $d$ results in
an increase in the kink velocity $v$
at the end points of the MT.
For realistic
biological systems, under normal circumstances,
the increase in $v$
can be even up to two orders of magnitude, resulting
in kink velocities
of
${\cal O}[10^2 m/sec]$~\cite{mtmodel}
at the end points of the MT.
\pr
As we shall argue now,
the parameter $d$
can be drastically affected
by an abrupt distortion of the
environment due to the influence of an
{\it external stimulus}. For instance,
it may be possible for
such abrupt distortions
to cause a local disturbance among the dimers,
so that
the value of $A$ is momentarily
diminished significantly, $d$ increases~\cite{mtmodel}, and the
condition
(\ref{reality})
is met.
However, such a distortion might affect the
order of magnitude estimates of the
effective mass scales involved in the problem, as we
discussed in section 2, and this may have consequences
for the decoherence time. So, we shall not consider it
for the purposes of this work.
More plausibly, an abrupt
environmental
distortion
will lead to a sudden increase in the
electric field $E$, which, in turn,
results in the
formation of a fast kink of $v \sim v_0$, via an increase in
$d$.
As we have seen above, this is not
unreasonable for the range of the parameters
pertaining to realistic biological systems.
In such cases the (fast) kinks are represented as
translational-invariant backgrounds of
non-critical string theories,
propagating with {\it real} velocities $v_s$ (\ref{qdef}).
This is the kind of structures that we shall mostly be
interested in for the purposes of this work.
Coupling them
to  quantum gravity will
lead to the collapse of the preconscious state,
as we shall discuss in the next
section.
\pr
The important advantage of formulating the MT system as
a $c=1$ string theory, lies in the possibility
of casting the friction problem in a Hamiltonian
form. To this end, we now make some comments on
the
various non-derivative terms in the
tachyon potential $V(T)\equiv U ( \psi ) $ in
the target-space effective action
\cite{banks}.
Such terms contributre higher order non-derivative terms
in the equations of motion (\ref{ten}). The term linear in $\psi $
is fixed in string theory and normalized (with respect to the
second derivative term ) as in (\ref{ten}) \cite{polch}.
The higher order  terms are polynomials in $\psi $ and
their coefficients can be varied according to
renormalization prescription \cite{banks}.
\pr
The general structure of the tachyon effective action in
the target space of the string is, therefore,
of the form \cite{banks}
\be
{\cal L} = e^{-\Phi }\sqrt{G} \left[ (T - c)^2 +
{\hat {\cal L}}(\nabla T , \nabla \Phi )\right]
\label{effact}
\ee
where $G _{\mu\nu}$ is a target space metric field,
and $c$ is a constant.
The linear term is the only universal term, showing
the impossibility of finding a stable tachyon background
in bosonic string theory, unless it is time dependent.
\pr
In our specific model, we have seeen that a term cubic in $\psi $
in the equation of motion (\ref{ten}), with a relative minus sign as compared to
the linear term, was responsible for the appearance of a kink-like classical
 solution.
Any change in the non-linear terms would obviously affect the
structure of the solution, and we should understand
the physical meaning of this in our biological system.
To this end, we consider a general polynomial in $T$
equation of motion for a static tachyon
in (1+1) string theory
\be
  T '' (\xi ) + \rho T' (\xi ) = P( T )
\label{polyn}
\ee
where $\xi $ is some space-like co-ordinate
and $P (T)$ is a polynomial of degree $n$, say.
The `friction' term $T '$ expresses
a Liouville derivative, since the effective
string theory of the displacement field $u$ is viewed as
a $c=1$
matter non-critical string. In our interpretation
of the Liouville field as a local scale on the world sheet
it is natural to assume that the single-derivative term
expresses the non-critical string $\beta$-function, and hence
is itself a polynomial  $R$ of degree $m$
\be
     T ' (\xi ) = R (T)
\label{secondpol}
\ee
Using Wilson's exact renormalization group scheme,
we may assume that $R(T) = a_2 T + a_4 T^2 $,
where  $a_2, a_4$ are related to the anomalous dimension
and operator product expansion coefficients for the tachyon
couplings.
The compatibility conditions for the existence of bounded
solutions to the equation (\ref{polyn}), then,
imply the form\cite{curiouseq}
$P(T) = A_1 + A_2 T + A_3 T^2 + A_4 T^4 $,
with $A_i = f_i (a_2, a_3), i =1,...4$.
Indeed such equations have been shown \cite{curiouseq}
to lead to kink-like solutions,
\be
  T (\xi ) = \frac{1}{2a_4} \{ {\rm sgn}(a_2a_4) a_2
{\rm tanh}[\frac{1}{2}a_2  (x-u t)]- a_2 \} \
\label{kink}
\ee
where the velocity $u=\frac{A_3 - 3a_2a_4}{a_4}$.
This fact
expresses for us
a sort of universal behaviour
for biological systems. This shows the existence of {\em at least}
one class of schemes which admit kink like solutions
of the same sort as the ones of Lal \cite{lal}
for energy transfer without dissipation in cells.
\pr
The importance of solutions of the form (\ref{kink}) lies in the
fact that they can be derived from a Hamiltonian and, thus, can be
quantized canonically \cite{hojman}.
They are connected to the solitons (\ref{eleven})
by a Renormalization Scheme change on the world-sheet of the
effective string theory, reproducing (\ref{ten}).
This
amounts to
the possibility of casting friction problems, due to the Liouville
terms, into a Hamiltonian form.
This is quite important for the quantization of the kink solution,
which will provide one with a concrete example of
a large-scale quantum coherent state for the preconscious
state of the mind.
In a pure field-theoretic setting, a quantization scheme
has been discussed in ref. \cite{tdva}, using
a variational approach by means of squeezed coherent
states. There is a vast literature on soliton
quantization using approximate $WKB$ methods.
We selected the method of ref. \cite{tdva}
for our purposes here,
because it yields more
accurate results than the usual $WKB$ methods
of soliton quantization\cite{WKB}, and it seems
more appropriate for our purposes here, due to its
direct link with coherent ground states.
We shall not give details on the derivation but
concentrate on the results. We refer the
interested reader to the literature\cite{tdva}.
A brief description of the method
is provided in Appendix B.
\pr
The result of such a quantization yields a
modified soliton equation for the (quantum corrected) field
$C(x,t)$ \cite{tdva}
\be
    \partial ^2 _t C(x.t) - \partial _x ^2 C(x,t)
+ {\cal M}^{(1)} [C(x,t)] = 0
\label{22c}
\ee
with the notation
\be
M^{(n)} = e^{\frac{1}{2}(G(x,x,t)-G_0(x,x))\frac{\partial ^2}
{\partial z^2}} U^{(n)}(z) |_{z=C(x,t)}
\qquad ; \qquad U^{(n)} \equiv d^n U/d z^n
\label{22}
\ee
Above, $U $ denotes the potential of the original soliton
Hamiltonian,
and $G (x,y,t)$ is a bilocal field that describes quantum corrections
due to the modified
boson field around the soliton.
The quantities $M^{(n)}$ carry information about the
quantum corrections, and in this sense the above scheme
is more accurate than
the WKB approximation \cite{tdva}.
The whole scheme may be thought of as a
mean-field-approach to quantum corrections to the soliton
solutions.
For the kink soliton (\ref{kink}) the quantum
corrections (\ref{22c})
have been calculated explicitly in ref. \cite{tdva},
thereby providing us with a  concrete
example of a large-scale quantum coherent state.
\pr
The above results on a consistent quantization of the soliton
solutions (\ref{kink}), derived from a Hamiltonian
function, find a much more general and simpler application
in our Liouville approach \cite{emninfl,emnest}.
To this end, we first note that
the structure of the equation
(\ref{polyn}), which leads to (\ref{kink}),
is generic for Liouville strings in non-trivial
background space-times.
If we view the Liouville mode as a local scale
of the renormalization group on the world-sheet \cite{emn,osborn},
one can easily show that for the coupling $g^i$ of any
non-marginal deformation $V_i$ of the $\sigma$-model
\footnote{For a concise review of the formalism and relevant notation
see Appendix A.},
the following
Liouville renormalization group equation holds
\cite{tseytl,emn}
\be
  {\ddot g}^i + Q {\dot g}^i = -\beta ^i = - G^{ij}
  \partial _i C [g]
\qquad ; \qquad Q^2 = \frac{C[g]-25}{6} + \dots
\label{st5}
\ee
where the dot denotes differentiation with respect to the
renormalization group Liouville scale $t$, and
$\dots $ denote terms removable by redefinitions
of the couplings $g^i$ (renormalization scheme changes).
The tensor $G^{ij}$ is an inverse metric in field space\cite{zam}.
Notice in (\ref{st5})
the r\^ole of the non-criticality of the string
($Q \ne 0$)
as providing a source of friction \cite{emn}
in the space of fields $g^i$.
The non-vanishing renormalization group $\beta$-functions
play the r\^ole of generalized forces.
The functional
$C[g]$ is the Zamolodchikov $c$-function, which is
constucted out of a particular combination
of components of the world-sheet stress  tensor
of the deformed $\sigma$-model \cite{zam,emn}.
\pr
The existence of friction terms in (\ref{st5})
implies a statistical description of the
temporal evolution of the system using
classical density matrices $\rho (g^i, t)$ \cite{emn}
\be
  \partial _t \rho  = -\{ \rho , H \}_{PB} +
\beta ^i G_{ij} \frac{\partial \rho }{\partial p_j}
\label{liouvmod}
\ee
where $p_i$ are conjugate momenta to $g^i$, and $G_{ij}\propto
<V_iV_j>$
is a metric in the space of fields $\{ g^i \}$ \cite{zam}.
The non-Hamiltonian term in (\ref{liouvmod})
leads to a violation of the Liouville theorem (\ref{one}) in the
classical phase  space $\{g^i, p_j \}$,
and constitutes the basis for a modified (dissipative)
quantum-mechanical
description of the system\cite{emn}, upon quantization.
\pr
In string theory, summation over world sheet surfaces
will imply quantum fluctuations of the string target-space
background fields $ g^i$\cite{emn,emninfl}.
{\it Canonical} quantization
in the space $\{ g^i \}$ can be achieved, given that
the necessary Helmholtz conditions \cite{hojman} can be shown to
hold in the string case \cite{emninfl}. The important feature
of the string-loop corrected (quantum)
conformal invariance conditions is that they can
be derived from a target-space action \cite{mm2},
which schematically can be represented as\cite{hojman,emninfl}
\bea
    {\cal S}  &=& -
    \int dt (\int _{0}^1 d\tau g^i E_i (t, \tau g, \tau {\dot g},
\tau {\ddot g} ) + total~derivatives \qquad ; \nn \\
 E_i (t,g,{\dot g},{\ddot g}) &\equiv &
 {\cal G}_{ij}( {\ddot g}^j
 + Q{\dot g}^i + \beta ^i )
\label{st6b}
\eea
where the tensor ${\cal G}_{ij}$ is a
(quantum) metric\cite{zam}
in theory space $\{ g ^i \}$. It
is characterized  by a specific behaviour \cite{emnest}
under the action of the renormalization group
operator ${\cal D } \equiv \partial _t + {\dot g}^i \partial _i$,
 \be
 {\cal D} {\cal G}_{ij} = Q {\cal G}_{ij} = <V_i V_j > + \dots
 \label{st6c}
\ee
where the $\dots $ denote diffeomorphism terms in $g$-space, that
can be removed
by an appropriate scheme choice\cite{emnest}.
\pr
In this way, a friction problem (\ref{st5})
can be mapped non-trivially onto a canonically
quantized Hamiltonian system, in similar spirit
to the solitonic point-like field theory
discussed in section 3.
The quantum version of (\ref{liouvmod})
reads \cite{emn}
\be
    \partial _t {\hat \rho} =
i [ {\hat \rho}, {\hat H} ] + i \beta ^i G_{ij} [ {\hat g}^i ,
{\hat \rho }]
\label{quantumliouv}
\ee
where the hat notation denotes quantum
operators, and appropriate quantum ordering is understood
(see below). We note that
the equation (\ref{quantumliouv})
implies that $\partial _t \rho$ dependes only on $\rho (t)$
and not on a particular decomposition
of $\rho (t)$  in the projections corresponding to various
results of a measurement process. This automatically
implies the absence of faster-then-light signals
during the evolution.
\pr
The analogy with the soliton case, discussed previously,
goes even
further if we recall \cite{emn} the fact that
energy is conserved in average
in our approach of time as a renormalization
scale.
Indeed, it can be shown that
the temporal change of the
energy functional of the string particle,
${\cal H}$, obeys the equation
\be
  \partial _t {\cal H} \propto {\cal D}<\Theta (z,{\overline z}),
\Theta (0) > = 0
\label{st6d}
\ee
where $\Theta $ denotes the trace of the world-sheet
stress tensor; the vanishing result is due to the
{\it renormalizability} of the $\sigma$-model on the
world-sheet surface, which, thus, replaces time-translation
invariance in target space.
\pr
However, the quantum energy fluctuations
$\delta E~\equiv~[<<{\cal H}^2>> - (<<{\cal H}>>)^2 ]^{\frac{1}{2}}$
are time-dependent :
\be
\partial _t (\delta E)^2
= -i <<[\beta ^i , {\cal H}]\beta^j G_{ji}>> =
<<\beta^j G_{ji} \frac{d}{d t}\beta ^i >>
\label{40}
\ee
Using the fact that $\beta ^i G_{ij} \beta ^j $ is a
renormalization-group invariant quantity,
we can express (\ref{40}) in the form
\be
  \partial _t (\delta E)^2  = -
<<Q^2  \beta ^i G_{ij} \beta ^j >>= -
<<Q^2  \partial _t C >>~\le 0
\label{fluct}
\ee
We, thus, observe that,
for $Q^2 > 0$ (supercritical strings), the
energy fluctuations decrease
with time for unitary string
$\sigma$-models \cite{zam}.
\pr
Before closing this section we wish to make
an important remark concerning
quantum ordering in (\ref{quantumliouv}).
The quantum ordering is chosen in such a way  so
that energy and probability conservation,
and positivity of the density matrix
are preserved in the quantum case. Taking into account
that in our string case
$\beta ^j G_{ij} = \sum _{n} C_{ij_1\dots j_n}g^{j_1} \dots g^{j_n} $,
with the expansion coefficients appropriate vertex operator
correlation functions \cite{emn},
it is straightforward to cast the
the above equation into a Lindblad form~\cite{lin}
\be
{\dot \rho } \equiv \partial _t \rho
= i [\rho, H] - \sum _m
\{ B^\dagger_m B_m, \rho  \}_{+}
+ 2 \sum _m  B_m\rho  B^\dagger _m
\label{bloch}
\ee
where the `environment' operators $B_m$, $B^\dagger_m$
are defined appropriately as `squared roots'
of the various partitions of the operator
$\beta ^j G_{ij} \dots g^i $.
This form may have important consequences
in the case one considers a wave-function
representation of the density matrix.
Indeed, as discussed in ref. \cite{gisin},
(\ref{bloch}) implies a {\it stochastic} diffusion
equation for a state vector, which has important
consequences for the localization of the wave-function
in a quantum
theory of measurement. We shall review briefly this
approach, and make a comparison with ours,
in the next section.
We stress, however,
that our approach based on density matrices
(\ref{bloch}) and renormalizability of the
string $\sigma$-model is more general
than any approach assuming state vectors.

\pr
\section{Quantum Gravity and
Breakdown of Coherence in the
String Picture of a Microtubule}

\subsection{Microscopic black holes
in MT chains as an environmental phenomenon}

\subsubsection{Formation of black holes}
\pr
Above we have established the conditions
under which a large-scale coherent state
appears in the MT network, which can be
considered responsible for loss-free energy
transfer along the tubulins.
Suppose now that {\it external stimuli}
produce  sufficient distortion in the
electric dipole moments of the water environment
of the MT.
As a result,
conformational (quantum) transitions
of the tubulin
dimers occur.
Such abrupt pulses
may cause sufficient distortion of the
space time
surrounding the tubulin dimers,
which in turn
leads to the formation of {\it virtual}
`black holes' in the effective target two-dimensional
space time.
Formally this is expressed by coupling the
$c=1$ string theory to two-dimensional quantum gravity.
This elevates the matter-gravity system to a
critical $c=26$ theory.
Such a coupling, then,
causes decoherence, due to induced instabilities
of the kink quantum-coherent `preconscious state',
in a way that we shall discuss below.
As
the required
collapse time of ${\cal O}(1\,{\rm sec})$ of
the wave function of the coherent
MT network is several orders of magnitude bigger than the
energy transfer
time  $t_T$ (\ref{transfer}),
the two mechanisms are compatible with each other.
Energy is transfered during the quantum-coherent
preconscious state, in $10^{-7} sec$,
and then collapse occurs to a certain (classical) conformational
configuration. In this way, Frohlich's frequency
associated with coherent `phonons' in biological
cells is recovered, but in a rather different setting.
\pr
We now proceed to describe the precise
mechanism for the breakdown of coherence,
once the system couples to quantum gravity.
To discuss this issue, we first note that there is
an exact conformal field theory \cite{witt},
a Wess-Zumino model over
$SL(2,R)/O(1,1)$ coset theory, whose target
space has the
metric of a two-dimensional black hole (to be defined below,
c.f. (\ref{st11})).
Its Euclideanized version is better studied,
and will be frequently used in this work, especially
when one studies non-perturbative configurations of
this field theory.
As a conformal field theory , Euclidean two-dimensional
black holes are described
by a Wess-Zumino
model over $SL(2,R)/U(1)$.
The presence of a black hole in the (1+1)-dimensional
effective string
model corresponding to (\ref{ten}) leads to
a non-trivial couping with target space-time gravity,
as well as the plethora of the $SL(2,R)$ global
$W_\infty$ states characterising two-dimensional
strings~\cite{emn}\footnote{It is
probably of interest to note that in two-dimensional
string theories, the target space time gravitons are
quasi-topological states, and as such the only non-trivial
geometries are either flrat space-times ($c=1$ string theory)
or the black hole
one, represented in ref. \cite{witt} as a coset
$SL(2,R)/O(1,1)$ Wess-Zumino model.
The physical states of these two models have been recently shown
to be equivalent~\cite{eguchi,ard}, thereby proving in a rigorous way
claims~\cite{witt,emn} that the $c=1$ flat string theory can be
considered as the asymptotic limit of the two-dimensional black
hole. It is also important to notice that,
for technical reasons, most of the computations and
arguments, especially in the context of the world-sheet
conformal field theory, refer to the Euclidean version
$SL(2,R)/U(1)$ of the black hole. Analytic continuation
is then assumed in order to transcribe the results in the
Minkowski case. Note, however, that
this procedure is not yet completely
understood in the context of string theory~\cite{verl}.}.
First, we discuss an explicit way of
dynamical creation of black holes in two-dimensional
string theory \cite{russo,witt}
through
collapse of tachyonic matter $T(x,t)$.
This is a procedure that can happen
as a result of quantum fluctuations
of various excitations.
\pr
Consider the equations of motion
for the graviton and dilaton fields obtained by
imposing conformal invariance in the model
(\ref{st2b}) to order $\alpha '$, ignoring the non-universal
tachyon terms in the potential.
It can be shown that a generic solution for the graviton
deformation
has the form \cite{russo}
\be
 ds^2 = -\{1 + \int ^x _{\infty} dx'[(\partial _{x'} T)^2
+ ({\dot T})^2 ] - \int ^t _{-\infty} dt' {\dot T} \partial _{x'} T \}
dt^2 + \{1 + \int ^t _{-\infty} dt' {\dot T}\partial _{x'} T \}dx^2
\label{st9}
\ee
Consider now an incoming localized wavepacket of the form
($a \equiv {\rm const}$)
\be
T = e^{-x} \frac{a}{cosh [2(x + t)]}
\label{st10}
\ee
It becomes clear from (\ref{st9}) that there will be
an horizon, obtained as a solution of the equation
derived
by imposing
the vanishing
of the coefficient of the $dt^2 $ term . Thus,
for late times $ t \rightarrow \infty$, the resulting
metric configuration
is a two-dimensional static black hole \cite{witt}
\be
      ds^2 = -(1 - \frac{4}{3} a^2 e^{-2x} ) dt^2
+ (1 - \frac{4}{3}a^2 e^{-2x} )^{-1} dx^2
\label{st11}
\ee
and the whole process describes a dynamical collapse of
matter.
The energy of the collapsing
wavepacket gives rise to the
Arnowitt-Deser-Misner (ADM)
mass $\frac{4}{3}a^2 $ of the black hole \cite{witt}.
It is crucial for the argument that there is no part
of the wave-packet reflected. Otherwise the resulting
ADM mass of the black hole will be the part of the
energy that was not reflected.
In realistic situations, the black hole is only {\it virtual},
since low-energy matter pulses are always reflected in two-dimensional
string theories, as suggested
by matrix-model
computations \cite{polchmatr}.
Indeed, if one discusses
pulses which undergo total reflection
\cite{russo},
\be
T (x,t) = e^{-x} \eta (x, t) \qquad ; \qquad
   \eta (x,t) =
   \frac{a}{cosh(2 (x + t))} +
   \frac{a}{cosh(2 (x - t))}
\label{refl}
\ee
then, it can be easily shown that there is
a {\it transitory} period where the space time geometry looks like a
black hole (\ref{st11}), but asymptotically in time
one recovers the linear dilaton (flat) vacuum\cite{aben}.
The
above example (\ref{refl})
gives a generic way of a (virtual)
dynamical matter
collapse in a two-dimensional stringy space-time,
of the type that we encounter in our model for the
brain, as a result
of abrupt quantum conformational changes of the
dimers.
In the case of MT,
the replacement $x \rightarrow \xi $, where $\xi $ is
the boosted coordinate in (\ref{eleven}),
is understood.

\subsubsection{(Infinite) Symmetry Breaking and Massless Modes}
\pr
Once a virtual black hole is formed in a MT chain,
the subsystem
of the displacement modes $\psi (\xi)$ becomes {\it open}
in a statistical mechanics sense.
This subsystem is a (1+1)-dimensional
non-critical (Liouville) theory. It is known that
the singularity structure of black holes in such
systems is described by
a topological $\sigma$-model,
obtained by twisting the $N=2$ supersymemtric
black hole~\cite{wbreak}. The field theory {\it
at the
singularity} is described by an enhanced
topological $W_{1+\infty} \otimes W_{1+\infty}$
symmetry. {\it Away} form the singularity,
such symmetry is {\it broken spontaneously }
down to a single $W_{1+\infty}$ symmetry,
as a result of the non-vanishing target-space gravitational
condensate~\cite{wbreak}. Spontaneous breaking in a
(1+1)-dimensional target string theory is allowed, in the sense that
the usual infrared infinities that prevented it
from happening in a local field theory setting are absent.
In ref. \cite{wbreak,emn} we have demonstrated
this phenomenon explicitly
by showing that, due to the (twisted)
$N=2$ supersymmetry associated with the topological
nature of the singularity, there is a suppression of the
tunneling effects, which in point-like theories
would prevent the phenomenon from occuring.
As a result,
the appearance of {\it massless}
states takes place~\footnote{The reader
might find curious the
appearance of {\it massless} excitations as a result of the
breaking of gauge
stringy symmetries, such as the target space
$W_\infty \otimes W_\infty $ in the two-dimensional
string theory~\cite{emn}. However, as we have mentioned
above and
in ref. \cite{wbreak}, these
symmetries are {\it topological}, and the massless
excitations are discrete non-propagating
states, which correspond to
longitudinal components
of string states that
cannot be gauged away~\cite{legpoles}.
This situation should be contrasted
to the case
of conventional gauge (local) symmetries in point-like
theories, where the massless Goldstone
excitation is `eaten up' (gauged away)
by the longitudinal component of the (propagating) gauge boson.
However, we mention that
even in the case of
point-like theories there are arguments for
the existence of a
Goldstone theorem for topological symmetries, such as the
flux symmetry in (2+1)-dimensional $QED$, whose massless
goldstone
particle is the photon itself~\cite{kovner}.},
which are delocalized global states, belonging to
the lowest-level of the string spectrum~\cite{wbreak}.
These are the leg-poles that appear in the scattering
amplitudes of $c=1$ Liouville theory~\cite{legpoles}.
Their excitation in the mind, results in
conscious perception, in a way similar to
the one argued in the context of local field theory
regarding
the excitation of the dipole wave quanta~\cite{delgiud}.
Here, however, as we shall see,
the mechanism of conscious perception
appears formally much more complicated, due to the
complicated nature of the enormous
stringy symmetries that are spontaneously
broken in this case.
.

\subsubsection{Evaporation of the microscopic Black Holes}
\pr
>From a formal point of view, the formation of virtual black holes,
with varying mass, can be modelled by the action of
world-sheet instanton deformations~\cite{emn}. The latter
have the property of shifting (renormalizing)  the
Wess-Zumino level parameter $k$ (related to central charge)
of the black hole conformal field theory of ref. \cite{witt}.
>From a conformal field theory point of view,
instantons are associated with induced {\it extra logarithmic}
divergencies (on the world sheet) in the presence of the
matter leg-poles. In our approach to target time, $t$, as a dynamical
world-sheet
renormalization group scale~\cite{emn}, $\phi = -t$,
such logarithmic divergencies, when
regularized, lead to extra
time dependences in the central charge of the theory, and
hence to a
time-dependent `effective' $k$, as mentioned above.
Some brief description of the formalism is given in Appendix A.
For mode details we refer the reader to the literature~\cite{emn,yung}.
It can be shown~\cite{emn} that
in such a case the $ADM$ mass of the black hole~\cite{adm}
depends on the scale (time)
\be
  M_{bh} \propto \frac{1}{\sqrt{k(t)-2}} e^a
\label{adm}
\ee
where $a$ is a constant that can be added to the dilaton field
without affecting the conformal invariance of the black hole solution
without matter. As shown in \cite{emn}, $k(t)$ actually increases
with time $t$, leading asymptotically to $\infty$, which corresponds
to the flat space-time limit. In that case, the system keeps `memory'
of the dilaton constant $a$, which pre-existed the black hole
formation. From a MT point of view, the constant $a$ corresponds
to a spontaneously chosen vacuum string state as a result
of spontaneous breaking mechanisms of, say, electric dipole
rotational symmetries etc,
in the ordered water
environment~\cite{delgiud}.
As we shall argue later on, this is important
for {\it coding} of memory states in this framework.
>From the above discussion it becomes clear
that {\it memory} operation in our approach is a two step
process: (i) formation of a black hole, of a fixed value
for the dilaton vacuum expectation value (vev)
$a$, related to spontaneous
breaking occuring in the water environment, as a result
of an {\it external stimulus}, and (ii) evaporation of
the black hole, due to {\it quantum} instabilities,
described by world sheet instanton effects in our
completely integrable approach to MT dynamics.
The latter effects drive the black hole
to a vanishing-mass limit by shifting $k$ and
{\it not } the dilaton vev, $a$. This implies
{\it storage} of information, according to
the general ideas of ref. \cite{emn}.
Indeed,
from a conformal field theory point of view, the constant $a$
can be shifted by exactly marginal deformations
(moduli) of the black hole $\sigma$ model, whose couplings
are {\it arbitrary}.
Such an operator has been constructed explicitly in ref.
\cite{chaudh},
and consists exclusively of combinations of global state deformations.
In the terminology of ref. \cite{chaudh} is called $L_0^2{\overline
L}_0$, and it is one of the (infinite number of) $W_\infty$ charges
that has been conjectured to characterize a two-dimensional
stringy black hole~\cite{emn}.
\pr

\subsection{Decoherence mechanism}
\pr
The important point to notice is that the
system of $T(\xi )$ coupled to a black hole
space-time (\ref{st11}),
even if the latter is a virtual configuration, it
cannot be critical
(conformal invariant) {\it non-perturbatively}
if the tachyon has a
travelling wave form. The factorisation
property of the world-sheet action  (\ref{st2b})
in the flat space-time case breaks down due to the
non-trivial graviton structure (\ref{st11}).
Then a travelling wave cannot be compatible with
conformal invariance, and renormalization scale dressing
appears necessary.
The gravity-matter system is viewed as a $c=26$ string
\cite{witt,emn}, and hence the
renormalization scale is {\it time-like} \cite{aben}.
This implies time dependence in $T (\xi, t)$.
\pr
A natural question arises whether there exist a deformation
that turns on the coupling $T (\xi )$ which is exactly
marginal so as to maintain conformal invariance.
The exaclty marginal
deformation of this black hole background
that turns on matter, $L_0^1{\overline L}_0^1$
in the notation of ref. \cite{chaudh},
couples necessarily
the propagating tachyon $T(\xi )$ zero modes to
an infinity of higher-level string states \cite{chaudh}.
The latter are classified according to
discrete representations
of the $SL(2,R)$ isospin, and together with the
propagating modes,
form
a target-space $W_\infty$-algebra \cite{emn,bakas}.
This coupling of massive and massless modes
is due to the non-vanishing Operator Product Expansion
(O.P.E.)
among the vertex operators of the
$SL(2,R)/U(1)$ theory \cite{chaudh}. The
possesses
an infinity of conserved charges in target space
\cite{emn} corresponding to the Cartan subalgebra
of the infinite-dimensional $W_\infty$ \cite{bakas}.
At present, an explicit construction~\cite{chaudh}
of the physical black hole moduli has been restricted
to the $L_0{\overline L}_0^2$
operator, mentioned above in connection with a
shif in the $ADM$ mass, as well as
the operator $L_0^1{\overline L}_0^1$ turning on matter
backgrounds. W
e believe, however, that the whole $W$-algebr              a
of charges corresponds to exactly marginal deformations
(moduli), leading to a quantum-coherence-preserving
moduli hair for the stringy black hole~\cite{emn}.
In practice, such global charges, which
contribute phase factors to the string universe wave function,
are impossible to measure by localized scattering
experiments in our world.
This, as we shall explain immediately below,
leads to the effective
breakdown of quantum coherence in the low-energy world~\cite{emn}.
>From the point of view of MT, such $W$ modes might be thought of as
constituting the
{\it `consciousness degrees of freedom'}~\cite{dn},
which in this picture, are not exotic, as suggested in ref. \cite{Page},
but exist already in a string
formalism, and they result in the {\it complete integrability }
of the two-dimensional black hole Wess-Zumino model.
\pr
This integrability persists quantization \cite{wu},
and it is very important for the quantum coherence
of the string black hole space-time\cite{emn}. Due to the
specific nature of the $W_\infty$ symmetries, there
is no information loss during a stringy black hole decay,
the latter being
associated
with
instabilities
induced by higher-genus effects on the world-sheet
\cite{emndec,emn}. The phase-space volume of the
effective field theory is preserved in time, {\it only
if} the infinite set of the global string modes
is taken into account. This is due to the string-level mixing
property of the $W_\infty$ - symmetries of the target space.
\pr
However,
any local operation of measurement, based on local scattering
of propagating matter, such as the functions performed
by the human brain, will necessarily break this coherence,
due to the truncation of the string deformation spectrum
to the localized propagating modes $T(\xi)$. The latter will, then,
constitute a subsystem
in interaction with an
{\it environment} of global string modes.
The quantum integrability of the full string system is crucial in
providing the necessary couplings.
This breaking of coherence
results in an arrow of time/Liouville scale,
in the way
explained briefly
above \cite{emn}. The black-hole $\sigma$-model
is viewed as a $c=26$ critical string, while the travelling
wave background is a non-conformal deformation.
To restore criticality one has to
dress $T(\xi )$ with a Liouville time dependence
$T(\xi, t)$ \cite{emn}.
>From a $\sigma$-model point of view,
to $O(\alpha ')$, a non-trivial consistency check
of this approach
for the black hole model of ref. \cite{witt} has been
provided in ref. \cite{emn}.
We stress, once more, that
the Liouville renormalization scale now is time-like,
in contrast with the previous string picture
of a $c=1$ matter string theory, representing the
displacement field $u$ alone before coupling to gravity.
\pr
By viewing the time $t$ as a local scale on
the world sheet, a natural identification
of $t$
will be with the logarithm of the area $ A $ of
the world sheet, at a fixed topology.
As the non-critical string runs towards the
infrared fixed point the area expands.
In our approach to
Liouville time
\cite{emn}, the actual flow of time
is opposite to the world sheet renormalization
group flow. This is
favoured by a bounce interpretation of the
Liouville flow due to specific
regularization properties of Liouville correlation
functions \cite{emn,kogan}.
This imlplies that we may set $t \propto -ln A$, with $A$
flowing always towards the infrared $A \rightarrow \infty$.
In this way, a {\bf Time Arrow} is implemented
automatically in our approach, without requiring the
imposition of time-asymmetric boundary conditions in
the analogue of the Hartle-Hawking state~\cite{hartle}.
In this respect,
our theory has many similarities to models of
conventional dissipative
systems, whose mathematical formalism \cite{dissip,santilli}
finds a natural application to our case.
\pr
With this in mind, one can examine the
properties of the
correlation functions of $V_i$, $A_N = <V_{i_1} \dots V_{i_N} >$,
and hence the issue of coherence breakdown.
In critical string theory such correlators
correspond to scattering amplitudes in the target-space
theory. It is, therefore, essential to check on
this interpretation in the present situation.
Since the correlators are mathematically
formulated on fixed area $A$
world-sheets, through the so-called fixed area constraint
formalism \cite{DDK,mm}, it is interesting to look for
possible $A$-dependences in their evolution. In such a
case their interpretation as target-space scattering amplitudes
would fail.
Indeed, it has been shown \cite{emnest}
that there is an induced target
space $A$-dependence of the regularized correlator $A_N$,
which, therefore, cannot be identified with a
target space $S$-matrix element, as was the
case of critical strings \cite{stringbook}.
Instead, one has non-factorisable contributions
to a superscattering amplitude $\nd{S} \ne S S^{\dagger} $,
as is usually the case in {\it open} quantum mechanical systems,
where the fundamental building blocks are density matrices
and not pure quantum states\cite{ehns}.
For completeness, we describe in Appendix A
some formal aspects of this situation,
based on results of our approach to non-critical
strings \cite{emn}.
\pr
It is this sort of coherence breakdown that we advocate as
happening
inside the part of the brain related to {\it consciousness},
whose operation is
described by the dynamics of (the quantum version of) the model
(\ref{five}).
The effective two-dimensional substructures, that we have
identified above as the basic elements
for the energy transfer in MT, provide the necessary
framework for coupling the
(integrable) stringy
black hole space-time (\ref{st11})
to the displacement field $u(x,t)$. This allows for
a qualitative description of
the effects of
quantum gravity on the
coherent superposition of the preconcious states.
\pr
At this stage some comments are in order concerning the
precise nature of the microscopic black holes.
It is not clear whether such black hole states
are related to
real (four-dimensional) quantum gravity effects
(e.g. spherically symmetric space-time singularities
), which couple to the quantum MT chains,
or they simply represent global environmental
effects in such systems\footnote{It should be
stressed that our environment-model-independent
stringy approach to the MT is applicable
even to cases when the virtual black hole creation in MT chains,
dicscussed above, represents not real quantum gravity effects,
but rather environmental entanglement with other biological
structures in the brain. In
such cases, the model (\ref{st2b}) is still valid,
but $\sqrt{\alpha '}$ is different from the Planck length, and
it is of order of the characteristic scale of the interactions
that cause
the
phenomenon.}.
In support of the former suggestion, we
now proceed in an estimate of the collapse time,
provided the latter is due to realistic
quantum gravity effects.
\pr

\subsection{Decoherence-time estimates }
\pr
One can calculate in this approach
the off-diagonal
elements of the density matrix in the string theory space $u^i$,
with now $u^i(t)$ representing the displacement field of the
$i$-th dimer. In a $\sigma$-model representation  (\ref{st2b}),
this is the tachyon deformation.
The computation proceeds analogously \cite{emn}
to the
Feynman-Vernon \cite{vernon}  and
Caldeira and Leggett \cite{cald}
model of environmental oscillators, using the influence functional
method, generalized properly to the string theory
space\footnote{Throughout this work, as well as
our previous works on the subject \cite{emn}, we assume that
such a space exists, and admits \cite{zam} a metric
${\cal G}_{ij}$. Indications that this is true are obtained
from perturbative calculations in  $\sigma$-model, to which
we base our belief.}
$u_i$. The
general theory of time as a world-sheet
scale predicts \cite{emn}
the following
expression for the reduced density
matrix\cite{vernon,cald} of the observable states:
\bea
\nonumber     \rho (u_I,u_F,  t  ) / \rho_S (u_I,u_F,  t  ) \simeq
e^{ - N \int _0^{t} d\tau \int_{\tau-\epsilon}^{\tau + \epsilon}
d\tau '
\frac{(u(\tau) - u(\tau '))^2}{(\tau - \tau ')^2}} \simeq \\
e^{- N \int_0^{t} d\tau \int_{\tau ' \simeq \tau }
d\tau '
 {\hat \beta} ^i G_{ij}(S_0){\hat  \beta} ^j }
 \simeq
e^{ - D N t ({\bf u_I}  - {\bf  u_F} )^2 + \dots }
\label{asym}
\eea
\nk where
the subscript ``S'' denotes quantities evaluated
in conventional
Schr\"odinger
quantum mechanics, and $N$ is the
number of the environment `atoms'\cite{emohn}
interacting with the background $u^i$.
For $(1,1)$ operators, that we are interested in,
the structure of the
renormalization group ${\hat \beta}$-functions
is
${\hat \beta}^i=\epsilon u^i + \beta^{ui} $,
where $\beta^{ui}=O(u^2)$ and
$\epsilon \rightarrow 0$ is the anomalous dimension.
Recalling \cite{emn}
the pole structure of the
Zamolodchikov metric,
$
G_{ij}=\frac{1}{\epsilon}{\cal G}_{ij}^{(1)} + regular$,
one finds that the dominant contribution
to the exponent
$K$
of the model (\ref{asym})
comes
from the
$\epsilon$-term in ${\hat \beta}^i $ and the
pole term in $G_{ij}$ :
\be
K=N \int _0^t {\hat \beta}^i G_{ij} {\hat \beta}^j d\tau \ni
2N\int _0^t u^i{\cal G}_{ij}^{(1)}\beta^{uj} d\tau + O(\epsilon )
\label{dcoeff}
\ee
\pr
Assuming slowly varying $u^i$ and $\beta ^i$
over the time $t$,
this implies
that the off-diagonal
elements of the density matrix
would decay exponentially to zero, within a collapse
time of order \cite{emohn}
\be
t_{coll} =\frac{1}{N} (O[ \beta ^{u^i} {\cal G}_{ij}^{(1)} u^j])^{-1}
t_s
\label{collapse}
\ee
in fundamental string units $t_s$ of time. The
superscript $(1)$ in the theory space metric
denotes the single residue in, say, dimensional
regularization on the world sheet \cite{emn,osborn};
$N$ is the number
of (coherent) tubulin dimers in interaction with the given
dimer that undergoes the abrupt conformational change.
Here the $\beta ^{u^i}$ -function is assumed to admit
a perturbative
expansion in powers of $\lambda _{s} ^2 \partial _{X}^2 $,
in target space, where the fundamental string unit of length
is defined as
\be
 \lambda _{s} =(\frac{\hbar \alpha '}{v_0^2})^{\frac{1}{2}}
\label{length}
\ee
where $v_0$ is the sound velocity (\ref{sound}), of order
$1 km/sec$ \cite{mtmodel}.
We work
in a system of
units where the light velocity is
$c=1$,
and
we use as the scale where
quantum gravity effects become important,
the grand unified string scale,
$M_{gus} = 10^{18} GeV$,
or in length $10^{-32} cm$,
which is $10$ times the conventional
Planck scale.
This is so, because our
model is supposed to be an effective description of quantum gravity
effects in a stringy (and not point-like)
space-time.
This scale corresponds to a time scale of $t_{gus}
= 10^{-42} sec$.
We now observe that, to leading order in the perturbative
$\beta$ function expansion in (\ref{collapse}), any
dependence on the velocity $v^2$ disappears
in favour of the scales $M_{gus}$ and $t_{gus}$.
It is, then, straightforward to
obtain a rough estimate for the collapse time
\be
    t_{col} =O[\frac{M_{gus}}{E^2N}]
\label{colltime}
\ee
where $E$ is  a typical energy scale in the problem.
Thus, we
estimate
that a collapse time
of ${\cal O}(1\,{\rm sec})$ is compatible with a number of coherent
tubulins of order
\be
 N \simeq 10^{12}
\label{number}
\ee
provided that the
energy stored in the kink background is of the
order of
$eV$. This
is indeed
the case of the (dominant) sum of binding  and
resonant transfer energies $\Delta \simeq 1 eV $ (\ref{energy})
at room temperature
in the
phenomenological model of ref. \cite{mtmodel}.
This number of tubulin dimers
corresponds to a fraction of $10^{-7}$ of the total
brain,
which is pretty close to the fraction believed
to be responsible for human perception on the basis of
completely independent biological methods.
\pr
An independent estimate
for the
collapse time $t_{col}$,
can be given
on the basis of
point-like
quantum gravity theory, assuming that the latter exists, either
{\it per se}, or as an approximation to some string theory.
One incorporates quantum gravity effects by employing
wormholes \cite{coleman} in the structure of space time, and then
applies the calculus of ref. \cite{emohn} to infer the
estimates
of the collapse
time. In that case, one evaluates the off-diagonal elements
of the density matrix in real configuration space $x$,
which should be compared to that in string theory space
(\ref{asym}).
The result of ref. \cite{emohn} for the time of collapse
induced by the interaction with a `measuring apparatus'
with ${\cal N}$ units is
\be
  t_{coll}' \simeq \frac{1}{{\cal N}} (M_{gus}/m)^3
  \frac{1}{m^3 (\delta x)^2 }
\label{moh}
\ee
where $m$ is a typical mass unit in the problem.
The fundamental unit of velocities in (\ref{moh})
is provided by the velocity of light $c$, since
the formula (\ref{moh}) refers to generic
four-dimensional
space-time effects. In the case of the tubulin dimers,
it is reasonable to assume that the pertinent
moving mass is the effective mass $M^*$ (\ref{effmass})
of the kink background (\ref{eleven}). This
will make contact with the microscopic model above.
 To be specific, (\ref{effmass}) gives
 $M^* \simeq 3 m_p$,
where $m_p$ is the proton mass.
This makes plausible the rather daring  assumption
that the nucleons (protons, neutrons) themselves
inside the protein
dimers are the most sensitive constituents
to the effects of quantum gravity. This is a
reasonable assumption if one takes into account
that the nucleons are
much heavier than the (conformational) electrons.
If true,
this would really imply, then, that elementary particle scales
come into play in brain functioning.
In this picture, then, ${\cal N} $
is the number of tubulin dimers,
and moreover in our case $\delta x = O[4 nm]$, since the relevant
displacement length in the problem is of order of the
longitudinal dimension of each conformational pocket in the
tubulin dimer.
Thus,
substituting these
in (\ref{moh})
one derives the result  ${\cal N} = N \simeq 10^{12}$,
for the number of coherent dimers
that induce a collapse within ${\cal O}(1\,{\rm sec})$.
\pr
It is remarkable that the final numbers agree
between these two estimates. If one takes into account
the distant methods involved in the derivation of the
collapse times
in the respective approaches, then one realizes that
this agreement
cannot be a coincidence. Our belief is that
it reflects the fundamental r\^ole
of quantum gravity in the brain function.
\pr
At this stage,
it is important to make some clarifying remarks
concerning the kind of collapse that we advocate
in the framework of non-critical string theory.
In the usual model of collapse due to quantum gravity \cite{emohn},
one obtains an estimate of the collapse of the off-diagonal elements
of the density matrix in configuration space, but no information
is given for the diagonal elements.
In the string theory framework
of ref. \cite{emn} the collapse (\ref{asym})
also refers to off-diagonal elements of the string
density matrix, but in this case the configuration space
is the string background field space.
In this case the off-diagonal element collapse suffices, because
it implies {\it localization} in string background space,
which means that the quantum string chooses to settle down in
one of the classical backgrounds, which
in the case at hand is the solitonic background discussed
in section 2.
This situation constitutes
the stringy analogue of the emergence of
almost classical states in an open quantum theory, as
a result of decoherence effects
induced by the environment~\cite{zurek}.
Such states, termed `pointer-states' in ref, \cite{zurek},
are minimum-entropy-producing states, which
behave almost classical under time evolution.
Due to the minimum-entropy production their time evolution
can be characterized as `almonst' reversible.
It should be stressed that this is a model dependent
statement, and the very existence of `pointer states'
depends crucially on the type of interaction with the
environment~\cite{albrecht}.
\pr
In ref. \cite{aspects}
we have discussed in detail the
emergence of `pointer states'
in a matrix-model
version of the model at hand, placed in a
random environment, assumed to simulate the quantum gravity
effects discussed in this section.
The configuration space of the inverted-harmonic oscillator
system, used to simulate the physics of the
matrix model, is actually a field space, since the
`position' variable $q$ of the model~\cite{matrix,aspects}
is a `collective co-ordinate'
related through a {\it canonical} transformation
to the `tachyon'
field (in our model this is the displacement field $u$).
It has been argued in ref. \cite{aspects}, based on known
results of the inverted-harmonic oscillator model~\cite{barton},
that the presence of an environment leads to the evolution of
a pure quantum state, represented as a Gaussian wave-packet,
at time $t=0$, into a probability distribution
for the particle, described by
the diagonal elements of the density matrix, which
retains the Gaussian shape but with a time- and temperature-
dependent width. The temperature indicates the effects
of the interaction of the system with a heat-bath, and as
discussed in ref. \cite{aspects} in our picture this
`temperature' is related to the deviations from the
conformal invariance of the system.
Such Gaussian states are distinct from Wigner coherent states,
which in the case of the inverted harmonic oscillator are not
square-integrable. These Gaussian wave-packets are
square integrable, and constitute
the
proper `pointer states' of the inverted  harmonic oscillator
problem\footnote{We note, for completeness, that
in the harmonic oscillator case of ref.
\cite{zurek} the pointer states are Wigner coherent states.}.
In our stringy interpretation of this system~\cite{matrix,aspects}
such `pointer states' are the closest analogues of classical
ground states of the string, which arise as a result
of quantum gravity environmental entanglement.
Of course,
the important question
`which specific background is
selected by the  above process ?' cannot be
answered unless a full string
field theory dynamics is developed. However, we believe that our
approach \cite{emn} of viewing the selection of a critical string
ground state as a generalized `measurement' process
in string theory space might prove advantageous over other dynamical
methods in this respect.
\pr
In the MT model, viewed as a completely integrable
non-critical
string theory, such classical ground states are the
solitons discussed in section 2.
The effects of quantum gravity in the one-dimensional
chain, amount to induce a collapse of (squeezed) coherent
states formed
around such soliton solutions (c.f. Appendix B), thereby
selecting a given classical ground state. This process is
interpreted in this work
as implying `conscious' perception.
\pr
An important question may arise at this point
concerning in connection with
the passage from quantum to classical worlds.
According to
a more conventional belief, the latter
should involve the collapse of the wave function
of the system. Indeed, from a formal point of view,
the density matrix refers to an ensemble of theories,
whilst the wave function is a quantity that characterizes
a single system. From the discussion above, we have shown how
decoherence (i.e. the collapse of the off-diagonal
elements of the density matrix in the space of string theories)
induces the appearance of clasical ground states of the
string, but we have not properly addressed the
question of how
a single ground is chosen.
As we mentioned previously, the question
`which ground state is going to be selected'
is a very difficult one and at present cannot be
answered without string field theory.
However, we can provide arguments in favour of
the localization of the state vector (if one insists
on using such a formalism) based on {\it specific }
propeties of our appoach to time as a Liouville field.
The most important of them is the {\it stochasticity}
that characterizes evolution in this  picture
to which we briefly turn now, for the benefit of the reader.
For more details of this approach we refer the interested
readers to the existing literature~\cite{emn,kogan}.

\subsection{Stochastic Nature of the
Renormalization Group Flow and State-Vector Localization}
\pr
To this end, we recall that in
Liouville theory a correlation function of $(1,1)$ matter
deformations $V_i$ is given by \cite{goulian}
\be
< V_{i_1} \dots V_{i_n} >_\mu = \Gamma (-s)
< V_{i_1} \dots V_{i_n} >_{\mu=0}
\label{Liouvcorr}
\ee
where $s$ is the sum of the appropriate Liouville energies,
and $<\dots >_\mu $ denotes a $\sigma$-model average
in the presence of an appropriate cosmological constant $\mu$
deformation on the world-sheet\footnote{In the case of
a black-hole coset model this operator is a
`modified cosmological constant' involving some mixing
with appropriate ghost fields parametrising the
$SL(2,R)$ string \cite{bershadsky}.}. The important
point for our discussion is the $\Gamma $-factor
$\Gamma (-s)$. For the interesting case of
matter scattering  off a two-dimensional ($s$-wave four-dimensional)
string
black hole, the latter
is excited to a `massive' (topological) string state
\cite{emnsel} corresponding to a positive integer
value for $s=n^+ \in {\bf Z}^+$.
In this case, the expression (\ref{Liouvcorr}) needs
regularization. By employing the `fixed area constraint'
\cite{DDK} one can use an integral representation for
$\Gamma (-s)$
\be
\Gamma (-s)=\int dA e^{-A} A^{-s-1}
\label{integralA}
\ee
where $A$ is the covariant area of the world-sheet. In the
case $s=n^+ \in {\bf Z}^+$ one can then employ a
regularization by analytic continuation,
replacing (\ref{integralA}) by a contour integral
as shown in fig. 4a \cite{kogan,emn}.
This is a well-known method of regularization
in conventional field theory, where integrals of
form similar to (\ref{integralA}) appear in terms of
Feynman parameters.
We note that it is the same regularization which was also
used to prove the equivalence
of the Bogolubov-Parasiuk-Hepp-Zimmerman renormalization prescription
to the dimensional regularization of `t Hooft \cite{BPZ}.
One result of such an analytic continuation is the
appearance of imaginary parts in the respective correlation functions,
which in our case are interpreted \cite{kogan,emn}
as renormalization group instabilities of the system.
\pr
Interpreting
the latter as an actual time flow,
we then may consider the contour of fig. 4a as implying
evolution of the world-sheet area in both
(negative and positive) directions of time
(c.f. fig. 4b), i.e.
\be
 Infrared ~ fixed ~ point  \rightarrow  Ultraviolet ~ fixed
~point \rightarrow
 Infrared ~ fixed ~ point
\label{flow}
\ee
In each half of the world-sheet diagram of fig. 4b
the Zamolodchikov $C$-theorem
tells us that we have an
irreversible Markov process.
According to the analysis of ref. \cite{emn}
the physical system is time-irreversible, since
the
physical processes associated with the
time transformations
in each `branch' of fig. 4b
are inequivalent. Indeed,
it has been argued in ref. \cite{emn} that
a highly-symmetrical phase of the two-dimensional
black hole occurs at the infrared fixed point of the
world-sheet renormalization group flow. At that point, the
associated $\sigma$-model is a topological theory
constructed
by twisting \cite{witt,eguchi} an appropriate $N=2$ supersymmetric
black-hole $SL(2,R)$ Wess-Zumino $\sigma$-model.
The singularity of a stringy black hole,
then, describes a topological degree of freedom.
The highly-symmetric phase is interpreted as
the state with the most `appropriate' initial conditions,
whose preparation requires finite entropy~\cite{emn}.
This in turn implies a `bounce' interpretation
of the renormalization group flow of fig. 4a, in which
the infrared fixed point is a `bounce' point,
similar to the corresponding picture
in point-like field theory.
Thus,
the ``physical'' flow of time
is taken to be {\it opposite} to the conventional
renormalization group flow, i.e. from the infrared
to the ultraviolet fixed point on the world sheet.
This can be confirmed explicitly,
by using world-sheet instanton calculus~\cite{emn}.
The instantons are semi-classical solutions,
whose existence can be established rigorously in
our black hole confrmal field theory~\cite{yung}.
In fact the black hole model is the limiting case that admits
(in its Euclideanized version) such solutions.
\pr
An important feature of this bounce picture
is that it allows for a {\it stochastic} evolution in time.
The quantum coupling constant space may  be
viewed as a stochastic manifold~\cite{emninfl}, where
the probability distribution   ${\cal P}(g^i, \tau |g^i_0)$
for the coupling/field  $g^i$   to evolve to this value
from the initial value $g_0^i$ at RG time $\tau=0$
obeys a Fokker-Planck equation (in the space of the
string backgrounds )
\be
\partial_\tau {\cal P}(\lambda, \tau) =
\frac{1}{8\pi^2}
\frac{\delta}{\delta \lambda ^i}
Q^{3}
\delta^{ij}
\frac{\delta}{\delta \lambda ^j}[Q^{3}
{\cal P} (\lambda,\tau)]
+ \frac{\delta }{\delta \lambda ^i}
[\beta ^i
{\cal P} (\lambda,\tau)]
\label{fp6}
\ee
modulo ordering ambiguities
for the $\lambda$-dependent diffusion coefficients.
$Q^2$ in the above formula denotes the `running'
(along RG trajectories~\cite{zam,emn})
central charge deficit of the effective theory.
In the case of our non-critical black hole string,
we identify $-\tau$ with the real target space
time $t$. The diffusion coefficient {\it decreases}
with time $t$, and vanishes
at the fixed point (equilibrium point)
thereby leading to {\it localization} in string theory
space (selection of a proper stable ground state as a result
of the dissipation induced by the non-vanishing $\beta^i$).
\pr
The stochastic equation (\ref{fp6})
plays an important r\^ole in
the connection of the collapse of
the off-diagonal elements of the density matrix,
discussed above, pertaining to an ensemble of theories,
and the {\it localization }
of the state vector of a (single) system.
Indeed, the Fokker-Planck equation for the
probability distribution cannot be derived
from the usual Schr\"odinger equation, which
is associated with a conservation equation
for the probability current rather, than a stochastic diffusion
equation.
As has been shown in ref. \cite{gisin},
if one insists on the existence of a
quantum state vector formalism, $\{ |\Psi > \}$,
defined
in the presence of an environment via the
density matrix $\rho(\Psi) ={\cal M} |\Psi ><\Psi |$,
with ${\cal M}$ a mean over the ensemble of theories,
then, the
environmental entanglement described by the
Lindblad equation (\ref{bloch}) corresponds to
a {\it stochastic} differential Ito process
for the state vector $|\Psi >$,
\bea
&~&|d\Psi > =-\frac{i}{\hbar} H |\Psi >
+ \sum _m (<B^\dagger _m >_{\Psi} B_m - \frac{1}{2}
B^\dagger _m B_m - \nn \\
-&~&\frac{1}{2} <B^\dagger _m>_{\Psi}
<B_m>_{\Psi })~|\Psi > dt +
\sum _m
(B_m - <B_m>_{\Psi})~|\Psi > d\xi _m
\label{ito}
\eea
where $H$ is the Hamiltonian of the system, $B^\dagger _m$,
$B_m$ are the environment operators,
$<\dots>_{\Psi}$ denote averages with respect to the
state vector $|\Psi >$, and $d\xi _m$ are complex differential
random matrices, associated with white noise Wiener or
Brownian processes~\cite{gisin}. In our completely integrable
non-critical string theory
representation of the MT system the environment operators
are related to the $\beta$-functions as explained previously
(c.f. (\ref{bloch})).
\pr
As regards eq. (\ref{ito}) attention is drawn
to a recent formal derivation~\cite{aremitsu} of this equation
from Umezawa's thermo-field dynamics model~\cite{umeztfd},
which is a quantum field theory model for open systems
interacting with the environment.
Note that,
in our interpretation of the time evolution as a
renormalization group evolution on the world sheet of the string,
white noise
distributions arise quite naturally
as a result of the summation
over world-sheet
topologies~\cite{friedan,emn,emninfl,wormhole,polchinski,paban,schmid2}.
Moreover, quantum-gravity entanglement
in such a framework may be described by a {\it canonically}
quantized formulation of string theory space~\cite{emninfl},
which makes the situation formally similar to thermo-field
dynamics~\cite{umeztfd}. However, there are important differences
in our approach, notably the absence of a doubling of the degrees
of freedom in order to simulate the environment, and also
the energy conservation on the average (\ref{st6d}),
as a result of
peculiar properties of the world-sheet renormalization
group~\cite{emn}.
A brief review of
the situation in the string case
is given in Appendix A.
\pr
Within the stochastic framework implied by (\ref{ito}) it
can, then, be shown
that under some not too resrtictive
assumptions for the Hamiltonian operator
of the system, localization of $|\Psi >$ within
a channel will always occur, as a result of
environmental entanglement.
To prove this
formally, one constructs a quantity that serves as a
`measure' for the delocalization
of the state vector, and examines its temporal
evolution. This is given by the so-called
`quantum dispersion entropy'~\cite{gisin}
defined as
\be
  {\cal K} \equiv -\sum _k <P_k>_{\Psi}ln~<P_k>_{\Psi}
\label{dispentr}
\ee
where $P_k$ is a projection operator in the `channel' $k$
of the state space of the system. Notice that from our point of view
the state space is the string background theory space.
This entropy is shown to decrease in situations where
(\ref{ito}) applies, under some
assumptions about the commutativity of the
Hamiltonian of the system with $P_k$~\cite{gisin}, which implies
that $H$ can always be written in a block diagonal form.
The result for the rate of change of the dispersion
entropy is then
\be
  \frac{d}{d t} ({\cal M}{\cal K}) = -\sum _k \frac{1 - <P_k>_{\Psi}}
{<P_k>_{\Psi}} R_k \le 0
\label{rate}
\ee
where $R_k$ are the (positive semi-definite)
{\it effective interaction rates} in channel $k$,
defined as~\cite{gisin}
\be
    R_k \equiv \sum _j |<L_{kj}>_{\Psi}|^2
\label{eir}
\ee
with $L_{kj} \equiv P_k L_j P_k $ denoting the projection
of the environment operators in a given channel.
>From a string theory point of view, the order of magnitude of the
environment operators
is that of the `square-root' of the
$\beta$-functions of the non-critical
string representing the system at hand (c.f. discussion
below (\ref{bloch})).
In the case of a system of $N$ test strings propagating
in non-trivial (non-critical) backgrounds $g^i$, then,
following ref. \cite{emohn}, one can argue that the strength of the
environmental entanglement is enhanced by $N$,
thereby leading to an enhancement of the effective interaction rate
(\ref{eir}) by $N$.
For the system of MT dimers at hand,
this
yields an estimate of the collapse-of-the-state-vector (or localization)
time of order
\be
t_{local} =
[\frac{{\cal K}(0)}{\sum _k <P_k>_{|\Psi>} (1-<P_k>_{|\Psi>})}]
\frac{1}{N (O[ \beta ^{u^i } {\cal G}_{ij}^{(1)} u^j])}t_s
\label{local}
\ee
in fundamental string units $t_s$ of time,
for a quantum mechanical network
of $N$ dimers interacting with the (gravitational) environment.
The above expression is rather formal.
We have made use of {\it normalized} non-singular interaction
rates~\cite{gisin}, $R_k/|<P_k>_{|\Psi>}|^2$, to estimate
the (non-critical-string) environmental gravitational entaglement,
and used the fact that the square of the environment operators
is of the order of the $\beta$-function of the non-critical string.
The probabilities $<P_k>_{|\Psi>}$ are (normalized)
probabilities
of selecting a particular background for the MT congifurations.
They depend on time $t$ (Liouville mode in our approach) through
the time-dependence of the state vectors $|\Psi > $.
They should be computed within a string-field theory framework,
and, hence, at present a detailed estimate of their magnitude is
impossible. However, one can make the {\it naturalness}
assumption that at $t=0$
\be
\frac{{\cal K}(0)}{\sum _k <P_k>_{|\Psi>} (1-<P_k>_{|\Psi>})} \simeq
O[1]
\label{orderone}
\ee
Taking into account that in string theory the square
of the $\beta$-functions decreases as the time flows (the string
approaches its conformal point and, therefore, the $\beta$ functions
tend to zero),
one observes that (\ref{local})
would yield an
estimate of the delocalization time of the form
form (\ref{collapse}).
This is the same
result as the density matrix collapse
we advocate in this work~\cite{emohn,emn}.
The similarity should have been expected from the fact
that equations (\ref{ito}) and (\ref{bloch}) are supposed
to describe the same physics.However,
we should stress, once more, that the above argument
made use of a naturalness assumption in order to infer
the order of magnitude of a combination
of probabilities of selecting certain backgrounds in
string field theory, which are quantities that at present
are impossible to be computed analytically.
\pr
Some comments are now in order, concerning the
precise application of the above localization-of-state-vector
theorem to our quantum gravity case.
In our case, quantum gravitational effects have non-zero
effective interaction rates because the latter represent the
recoil of the space-time itself under matter scattering off an
effective
microscopic virtual black hole formed in the
MT chains, as a result of abrupt conformationl changes of
the dimers. The latter phenomenon
can be a result of an {\it external stimulus},
including realistic four-dimensional quantum gravity fluctuations.
The very nature of the microscopic black hole, of having $ADM$
mass of order of $M_{Planck}$, which, then, evaporates
in the foam,
makes it impossible to be treated
semi-classically, by adding a hamiltonian term in the effective
action. Hence, in the presence of such a black hole
coupling of an MT chain
to `environmental' operators appears necessary,
and the effective interaction rate $R_k$ is non zero, thereby
producing the localization effects on the state vector
of the system.
\pr
The effective interaction rate
vanishes when a {\it stable ground state} is produced,
as is the case of dissipation (friction).
Our stringy system exhibits `gravitational
friction'~\cite{emn}. The permanent
ground state is the critical string vacuum.
Then localization stops. This is exactly the situation
encountered
in the {\it `pointer states'} approach,
discussed in ref. \cite{zurek}
and, in connection with the present problem, in ref. \cite{aspects}.
It is our point of view that the `pointer state' approach, via
decoherence, is more general and probably more suitable
for describing the actual situation encountered in
`conscious perception' by the brain. The latter is associated
with the emergence of {\it almost classical} states as a result
of {\it decoherence}, which evolve in time
almost as classical objects, thereby realizing -
for
all practical purposes-
the
transition from quantum to classical worlds.
The estimated time of such a decoherence is argued in this
work to be of
${\cal O}(1 sec)$. As we have discussed above, if one insists on using
state vectors, which may not be necessary, then
the localization time of the state vector
can be of the same order, although the latter result relies
on some naturalness assumptions in string field theory.
We should stress that the state vector in this case {\it does not}
satisfy a simple Schroedinger equation due to the environmental
effects, which convert the latter to a stochastic equation
of Ito type (\ref{ito})~\cite{gisin}. This is
a formal difference from the approach of ref.
\cite{zurek}, but all the ideas in that work, concerning
the effects of decoherence in the transition from the quantum
to classical,
are perfectly
consistent with our approach here, and also in ref. \cite{emohn,emn}.
\pr
The stochastic framework described above, which is inherent in our
Liouville approach to the concept of time, leads also to
some other interesting properties of ensembles of one-dimensional
systems, which might be of relevance to properties of MT
networks that have been observed experiumentally.
We now briefly turn to some of them.

\subsection{Possible connection with conformal models
of disorder}
\pr
The appearance of extra logarithmic divergencies on the
world sheet as a result of instantons, discussed above and
in Appendix A,
which cannot be accounted for by a standard renormalization
group flow of a given $\sigma$-model, has been interpreted
in ref. \cite{emn} as expressing a {\it change}
in conformal field theory describing matter back-reaction effects
on the space-time metric as the
string propagates
in a black hole background.
\pr
What we shall argue below is that,
from a formal point of view, such effects may be thought of as being
associated with
certain operators of the underlying conformal field theory
with zero conformal dimension, which
play a non-trivial r\^ole in a black hole background.
Such a situation has been recently argued to be
the case of certain conformal models of disorder
in condensed matter physics~\cite{tsvelik}.
Indeed, it has been shown in ref. \cite{tsvelik}
that in certain ($1+1$)-dimensional conformal
models with disorder
there are operators with degenerating
conformal dimension
in the so-called replica limit (employed to deal
effectively with disordered systems). In this limit
the central charge of the theory vanishes.
Such operators
dissapear from the spectrum of the theory in the O.P.E.
of two primary field operators of the conformal model,
but they leave behind unexpected logarithmic divergencies
in correlation functions, which accompany extra operators
termed {\it logarithmic operators}~\cite{tsvelik}.
\pr
If we view the black hole model as a completely integrable
critical string ($c_{tot}=26$), then including the
ghost contrbutions arising from a gauge-fixing of the world-sheet
reparametrization invariance~\cite{stringbook}, the total
central of the theory vanishes, and the theory is conformal.
The
Hamiltonian of the string is given by the world-sheet Virasoro
operator $L_0$. The conformal weight (dimension), $h _\alpha$,
of a string state
$|{\cal E} _\alpha >$
is defined as
\be
    L_0 |{\cal E} _\alpha > = h _\alpha  |{\cal E} _\alpha >
\label{confweight}
\ee
and similarly for the antiholomorphic part (denoted by a bar).
\pr
Given that
effects
associated with a
change in conformal field theory,
such as back-reaction of matter on the structure of space-time etc,
are purely stringy effects,
the most natural way of incorporated them in a $\sigma$-model
language is to go {\it beyong a fixed genus} $\sigma$-model
and consider the effects of resummation over world-sheet
genera. This is a very difficult procedure to be carried out
analytically. However,
for our purposes a sufficient analysis,
which describes the situation satisfactorily (at least at a
qualitative level), is that of resumming one-loop (torus)
world-sheets. Non-trivial effects arise from degenerate
Riemann surfaces, namely from long-thin world-sheet tubes
that are attached to a Riemann surface of lower genus (sphere
in this case). Such effects are similar to `wormholes' in a
four-dimensional quantum gravity treatment~\cite{wormhole}.
\pr
>From a formal point of view, such insertions in a Riemann
sphere are described
by adding bilocal world-sheet operators  on the
world-sheet sphere~\cite{polchinski}
\be
 {\cal B}_{ii} =\int d^2z V_i (z) \int d^2w V_i (w)
 \frac{1}{L_0+{\overline L}_0 -2}
\label{bilocal}
\ee
where the last factor represents the string propagator
on the world-sheet cylinder, with the $L$'s denoting
Virasoro operators.
By inserting a
complete set of intermediate string states~\cite{polchinski},
${\cal E}_\alpha$,
we can express the world-sheet `tube' in
(\ref{bilocal}) as an integral
over the radius $q$ of the tube's cross section
\be
  \sum _\alpha   \int dq d{\overline q}
  \frac{1}{q^{1-h_\alpha} {\overline q}^{1-{\overline h}_\alpha}}
\{{\cal E}_\alpha (z_1)
\otimes (ghosts) \otimes
{\cal E}_\alpha (z_2) \}_{\Sigma \oplus \Sigma '}
\label{props}
\ee
where $h_\alpha, {\overline h}_\alpha $ are conformal dimensions
of the states/fields ${\cal E}_\alpha$,
i.e. $L_0{\cal E}_\alpha =h_i {\cal E}_\alpha $.
The sum in (\ref{props}) is over
states propagating along the thin tube, connecting
the world-sheet pieces $\Sigma$ and $\Sigma '$ (in the case
of the degenerating torus handle of interest to us,
$\Sigma = \Sigma '$).
The terms `ghosts' indicate
appropriate insertions of ghost fields.
This has to do with the fact
that the total central charge of the ghost and
matter theory vanishes in the critical string model of
ref. \cite{polchinski}.
\pr
One easily observes that
{\it extra} logarithmic divergencies in (\ref{props}),
not included in the perturbative $\beta$-function
analysis on $\Sigma$, and thereby capable of expressing
quantum effects associated with a change in conformal field
theories,
may come from
states with $h_\alpha ={\overline h}_\alpha =0$.
If such states are {\it discrete} in the space of states, i.e.
they bear non-trivial contributions
to the sum-over-states
in (\ref{props}), then
will
lead to divergencies
in (\ref{bilocal}).
These properties are obviously
{\it background dependent}.
\pr
The bilocal term (\ref{bilocal}) can be writen as a local
world-sheet effective action
term, if one employs the well-known trick
of wormhole calculus~\cite{wormhole}
by writing
\be
  e^{{\cal B}_{ii}} \propto \int d\alpha
  e^{\alpha _i^2 + \alpha _i\int V_i }
\label{wlocal}
\ee
and this results in a `quantization' of the theory space
$\{ g^i \}$ of the string. The requirement for a canonical
quantization are met by the non-critical (Liouville strings)
as discusssed in detail in ref. \cite{emninfl}.
The highly non-trivial point we wish to make is that
in case non-trivial logarithmic divergencies appear
their effect in the quantized renormalization-group flow
will be to induce, by their absorption,
renormalization-group-scale-dependent `widths' in the
distribution functions of families of theories
(Gaussian wave-packets) in
quantum theory space~\cite{schmid2}. Notably such distribution
functions, with widhts that are
time-dependent (since in our approach
target-time is a quantum RG scale),
have arisen naturally as `pointer' states~\cite{aspects}
in our (world-sheet-genus resummed)
matrix-model approach~\cite{matrix}
to the black hole problem. As we discussed above, such states
played an important r\^ole in the transition from the
quantum to the classical world, and hence to the
concious perception of the brain.
\pr
To understand better the physical situation,
and support the conjecture that
the black hole model has such zero modes, we compare
our system to that
of
(critical) string theories propagating in non-trivial
backgrounds of extended objects in target space (e.g.
solitons, instantons etc), where the appearance of such
non-trivial
zero modes
appears to be a generic feature~\cite{paban}.
For instance,
consider the case of matter (particle) scattering
off a monopole background in the target space of
the string~\cite{paban}.
In this case, the recoil of the scatterer (monopole)
cannot be described by a single conformal field theory,
but by a change in conformal field theories (this is similar
to our back-reaction in the black hole case).
In
the example of ref. \cite{paban}, where one linearizes
around a target-space monopole background,
there are
(BRST-null) zero modes in the spectrum of $L_0$,
associated with the position
of the monopole center of mass in target space, which
breaks translational invariance.
The relevant marginal operators therefore
are constructed appropriately out of Noether currents.
Such operators are total derivatives on the world sheet
($\theta$-terms)
  \be
    V_k \propto \partial _\alpha J_k ^\alpha \otimes~(ghosts)
\label{noether}
\ee
where the
ghost insertions ensure
vanishing conformal dimension of $V_k$, and
$J_k^\alpha$
(with $\alpha$
a world-sheet
index)
is a target-space
translation Noether current,
represented as a world-sheet object,
corresponding to the translation
\be
   X^k \rightarrow X^k + u^k
\label{transl}
\ee
Circulation of the `zero modes' (\ref{noether})
along the long tubes (\ref{props}) of
pinched world-sheet surfaces, then, produces logarithmic
divergencies~\cite{paban}.
Restoration of conformal invariance
induces a recoil of the monopole so that the total
target momentum of the system `particle $+$ monopole (scatterer)'
is preserved. This expresses
the physical process of
the recoil of the monopole
scatterer,
which in this way
cancels the divergent effects appearing
in
the scattering of the particle off the monopole.
\pr
A similar thing happens in the black hole case~\cite{emn}.
Scattering of propagating matter off the black hole
results in a change of state of the black hole background~\cite{emnsel}.
Such a change cannot be described by a single conformal field theory,
because it is a quantum effect in target space.
In the black hole case at hand
there are conformal-dimension zero
operators in the spectrum, associated with the $W_\infty$
global modes~\cite{chaudh,emn}. Such modes express the `recoil'
of the black hole scattering center, during matter
propagation (scattering) in a black hole background.
Propagation of such modes along the thin tubes of degenerate
Rieman surfaces (\ref{props}) will, then, lead to
divergencies.
\pr
Semi-classically such effects can be represented qualitatively
by the world-sheet
instantons, which in this way represent higher-genus as well as
global mode effects~\cite{emn}.
Especially in the topological (twisted supersymmetric)
model of the black hole singularity~\cite{emn}, such instanton
deformations may be expressed as total derivative
terms in moduli space, and in particular as BRST $Q$-exact
states~\cite{yung,emn}, where $Q$ is the BRST charge in a string
language\footnote{The ghost fields in this model originate from the
fermion fields of the $N=2$ supersymmetry whose twisted version
produces zero central charge~\cite{yung,emn} For more details
on the construction, as well as properties of the instanton,
anti-instanton deformations we refer the reader to the
literature~\cite{yung,emn}.}. This total-derivative $Q$-exact form
of the instanton deformations
in the black hole model, describing `recoil'/back-reaction
of the space-time itself,
is suggestive enough to look for an
analogy with the
monopole-string case of
ref. \cite{paban}
discussed above.
It is worthy of pointing out that the similarity of the
black hole problem to that of a string soliton is not
unrelated to the connection between such systems
as a result of (non-perturbative)  {\it duality}
symmetries~\cite{strom}.
A more detailed study of these considerations is
in progress~\cite{emnest}.
\pr
Contact with conformal field theories of disorder is, therefore,
made
upon the observation that
the presence of operators
of zero conformal dimension
in string models
in non-trivial space-time backgrounds,
whose propagation
along degenerate handles produce extra logarithmic
divergencies,
also characterizes certain models
of disorder in condensed matter physics~\cite{tsvelik}.
Obviously such operators in the disordered systems
are similar to the zero modes
of the soliton or black-hole string,
mentioned above. Circulation
of such operators
along degenerate Riemann
tubes (\ref{props}) (wormholes)
will produce logarithmic divergencies not accounted for by the
conventional Renormalization Group flow at a fixed genus.
In
our opinion,
considering the effects of logarithmic operators
in resummed world-sheet
might also have
relevance
for the theory of disorder itself,
as
describing the recoil of the impurities.
>From a target-space quantum gravity point of view,
disorder may exhibit itself as a vanishing
vacuum-expectation-value of the metric tensor
in target space $< G_{MN} >=0$, which would be
a topological (highly-symmetrical)
phase of string theory (`disordered' phase). As the energy goes down,
diffeomorphism invariance is broken and a non-zero
$<G_{MN}>$ is achieved (`ordered' phase).
As we discussed in ref. \cite{emn}, and mentioned above,
a situation like this
has been argued to arise in the topological $N=2$ $\sigma$-model
that arguably simulates
the singularity of the black hole~\cite{wbreak}. This may be
interpreted as
implying
that formation
of black holes contributes to disorder.
Subtleties concerning the precise meaning of
an order parameter in our two-dimensional target space time,
of relevance to our present approach to MT dynamics,
have been discussed in ref. \cite{wbreak} and are related
to a Kosterlitz-Thouless (or `topological') order/disorder
phase transition in this case.
Of course all these are speculations.
To fully understand the phase transition
one has to resort to string field theory, or to a complete theory
of quantum gravity, something which at present is not possible.
\pr
>From our point of view in this work,
a connection of our quantum-gravity situation
with the conformal field theory of
disorder~\cite{tsvelik} as a conformal-field-theory changing
situation,
if formally established, will be a very
important step towards a mathematically rigorous formulation
of the process of conscious perception as a
`disorder/order' process. The r\^ole of the external stimuli
inducing microscopic black hole formation, and thereby
disorder (as a back-reaction/recoil process) in the MT chains,
may, then, stand a good chance of being
formulated in a rigorous way, by exploiting the conformal
structure of our stringy MT model, stemming from its
complete integrability. We believe that our work in this article
provides sufficient
motivation for
such studies, either in the continuum formalism of our model
of MT, or preferably
in the truly non-perturbative
lattice (random surface)-numerical simulation approach,
using triangulations of the Euclidean
world-sheet surfaces employed
in the model. In connection with the latter analysis, it is
worthy of mentioning that our approach to target time, viewing the
latter as
a local dynamical (Liouville)
scale on the world sheet,
employs Euclidean world-sheets, and the target space
Minkowski signature is an induced phenomenon
in the supercritical {\it
effective} string model ($C > 25$) of the propagating matter in
an evaporating black hole background~\cite{aben,emn}.
This is essential for numerical simulations, which
employ
triangulations of
euclidean world-sheets.
We
hope to turn to these issues
in the near future.

\section{Possible Experiments to check on the above ideas}
\pr
It is the purpose of this section
to consider possible applications of the above formalism
of
MT as (completely integrable) non-critical (Liouville) strings
in explaining or predicting properties of these systems
under some situations that can be met experimentally
in the laboratory.

\subsection{Growth (Dynamical Instability)
of a Microtubule Network and Liouville Theory}
\pr
The above considerations are valid
for MT networks whose size of individual MT
is larger than a certain critical size \cite{mtmodel}.
Kink-like excitations, that were argued to be crucial
for the physics of the conscious functions of the brain,
cannot form for small microtubules.
The question, therefore, arises whether the non-critical
effctive string theory framework described above
is adequate for describing the growth process
associated with the formation of a MT network.
This phenomenon is physically  and biologically very
interesting since these structures are known
to be the only ones so far that exhibit
the so-called `{\it dynamical instability growth}' \cite{mitchison}.
This is an out-of-equilibrium process acording to which
an individual MT can switch randomly between an
`assembly state' (+), in which the MT grows with a
speed $v_+$,
and `diassembly state' (-), in which the MT shrinks
with velocity $v_{-}$.
Recently there have been attempts to construct
simple one-dimensional
theoretical models with diffusion~\cite{growth}
that can describe qualitatively the above phenomenon.
An interesting feature of these models, relevant to our framework,
is that for a certain range of their parameters exhibit
a phase transition to an unbounded growth
state\footnote{This situation may also be viewed from a
spontaneous-symmetry breaking point of view
along the lines of ref.
\cite{ssb}, which in one dimension can occur
when the system is out of equilibrium.
This second point of view is more relevant to
our non-critical string approach, which as we
explained earlier is an out-of-equilibrium
process.
The symmetry breaking can be exhibited
easily by looking at one-dimensional models with driven diffusion
of say two species of particles corresponding
to the (+) and (-) conformational
states of the MT growth.
In the broken phase there is a difference between the
`currents' correpsonding to the (+) or (-) states.
Under certain plaussible conditions \cite{ssb},
associated with formation of droplets of the
`wrong sign' in any given configuration of the
above states,
the system can
switch between (+) and (-) states
with a switching time that depends on
dynamical parameters. For a certain range of
these parameters the switching time is of order
$e^{N}$, where $N$ is the size of the system, thereby
implying spontaneous symmetry breaking
in the thermodynamic limit where $N \rightarrow  \infty$.}.
In the case of MT networks, it is known experimentally
that a `sawtooth' behaviour in the time-dependence
of the size of a MT (Fig. 3),
occurs as a result of hydrolysis
of GTP nulceotides bound to the tubulin proteins
(i.e. the transformation GTP  $\rightarrow$ GDP )\cite{mitchison}.
The hydrolysis is responsible for providing the
necessary free energy for the conformational changes of
the tubulin dimers: the dynamical instability phenomenon
pertains to polymerization of the GTP tubulin, while
the GDP tubulin stays essentially unpolymerized.
In view of quantum oscillations between the
two conformations of the tubulin, and the above
different behaviour of polymerization,
it is natural to conjecture that quantum
effects may play a r\^ole in the MT growth process.
\pr
Thus, our Liouville theory representation
of the effective degrees of freedom $u$ involved
in the model of \cite{mtmodel}, invented to explain
classical aspects of the hydrolysis of GTP $\rightarrow$ GDP,
might, in principle, be able of explaining qualitatively
the `sawtooth' behaviour of Fig. 3, even before the formation
of kinks.
To this end, we remark
that in Liouville dynamics, with the Liouville
scale identified with the target time \cite{emn},
the inherent non-unitarity (in the world-sheet)
of the Liouville mode implies that
the central charge $C$ of the theory flows
with the scale in such a way that
near fixed points it oscillates a bit
before settling down. Indeed,
for a non-critical string with running
central charge $C[g, t]$ , $t$ is the Liouville scale/time,
the following second order equation (local in target space-time)
holds near a fixed point of the Renormalization Group Flow~\cite{tseytl}
\be
 {\ddot C}[g, t]
 + Q [g,t] {\dot C} [g,t] \le 0 ~for C \ge 25
 \qquad ; \qquad  Q^2 [g, t] =\frac{1}{3}
(C[g,t ] - 25)
\label{ctheorem}
\ee
This is a local phenomenon in target space-time.
Globally,
there is a preferred direction
in time along which the
entropy of the system increases \cite{zam,emn}.
\pr
The small oscillations of $C$
in (\ref{ctheorem})
may be attributed to the double direction
of Liouville time that arises as a result
of imaginary parts (dynamical instabilities)
appearing due to
world-sheet
regularization by analytic continuation
of non-critical string correlation functions
\cite{kogan,emn} (Fig. 4)
This point is discussed briefly in Appendix A.
\pr
Along each direction of Liouville time in Fig. 4
there is an associated variation of $Q [g,t]$ and $C[g,t ]$,
and the volume of the `one-dimensional universe' of MT
either increases or decreases \cite{emninfl}. Thus,
as a result of the oscillations of the Liouville dynamics
(\ref{ctheorem}) a `sawtooth' behaviour of MT, which expresses the
result
of polymerization of GTP tubulin,
can be qualitatively
explained within the framework of non-critical Liouville dynamics.
The analogy of the
above situation
to a stochastic
expansion of the Universe (inflation)
in non-critical string theory as discussed in ref.
\cite{emninfl} should be pointed out.
\pr
It should be noted at this stage that
in this effective framework the origin of non-criticality
of the subsystem of tubulin dimers is left unspecified.
Quantum Gravity fluctuations appear on an equal footing
with the environment of the nucleating solvant that
surrounds the MT in their physical environment.
The distinction can be made once a detailed
description
of the environment is given, which of course
would specify the form of the target metric
coupled to
the matter system of tubulins. As we have discussed in previous
sections,
in the quantum Gravity
case this is achieved by the
exactly marginal operators of the $SL(2,R)/U(1)$
conformal field theory that describes
space-time singularities in one-dimensional strings \cite{witt}.
The latter involve global (non-propagating)
string modes which cannot be detected
by localized scattering experiments \cite{emn}.
On the other hand, it seems likely that
an exact conformal field theory
that describes the nucleating solvant does not exist.
However, the effective Liouville string that describes
the embedding of a MT in it, and the associated
environmental entanglement, is obtained by a
simple Liouvlle dressing of the model discussed in section 3.
The information about the environment is hidden in the
form of the target metric that couples to the system.
The fact
that {\it both} the formation of an MT via the dynamical instability
phenomenon,
and the quantum
gravity effects on MT, involve conformational changes of the tubulin
supports the above point of view.
Of course the strength of the dynamical instability
in case the latter is due to quantum gravity fluctuations
will be much more suppressed as compared to the hydrolysis case.
In that case,
there will not be sufficient time
for complete polymerization
of the GTP tubulin conformation. In such a case
one would probably expect `sudden'
`sawtooth' peaks of the tip of the MT
whose magnitude will be affected by the order of
quantum gravity entanglement.
The growth in this case will be bounded.
\pr
It should be mentioned that recently there have been
some
experiments claiming such length fluctuations
\cite{kkk}, in the case of
carbon nanotubes. The authors of ref. \cite{kkk}
claimed that they observed such changes in the length
of the tubes, which they attributed to sudden jumps
of the respective wavefunctions, according to
the approach of Ghirardi Rimini and Weber \cite{grw}.
In this respect, one should think of repeating the tests
but with isolated MT\cite{rosu}.
We should stress, however, that the above discussion is at this stage
highly speculative and even controversial, given that
there appear to be conventional explanations
of this phenomenon in carbon nanotubes\cite{conv}.
Of course such conventional explanations
do not exclude the possibility of future observations of
quantum-gravity
induced bounded growth in MT,
along the lines
sketched above.
\pr
We close this section by mentioning that
in realistic situations the growth process of
an MT network is not unlimited.
After a critical length  is
exceeded, the growth is saturated and eventually stops.
The formation of kink excitations (\ref{eleven})
might be important for this
auto-regulation of the MT growth \cite{mtmodel}.
For more details on the conjectural r\^ole of the kinks
in this growth-control mechanism we refer the reader
to the literature \cite{mtmodel,growth}.
\pr
The above situation
should be compared with
the limitations on
the (stochastic) Universe expansion
in the non-critical-string-driven inflationary
scenario of ref. \cite{emninfl}.
There, it can be shown that within our framework
of identifying the Liouville field with the target time,
the average density
$<<\delta _s>> \equiv Tr(\rho \delta _s ) $
of the
non-critical strings
that drive the inflationary scenario
obey an equation of the form \cite{turok,emninfl}
\be
         \partial _t <<\delta _s >> = -aQ<<\delta _s >>
+ bQ^3 <<\delta _s >>
\label{denst}
\ee
where $\alpha $,$b$ are positive quantities,
computable in principle in
the Liouville-string framework \cite{emninfl}.
The first term in (\ref{denst})
is due to the (exponential) expansion
of the string-Universe volume, and the second term
corresponds to the regeneration of strings
via breaking of large strings whose size exceeds that of
the Hubble horizon \cite{emninfl}.
This second term comes from the diffusion due to the
non-quantum mechanical terms in the equation
(\ref{quantumliouv}).
At early times the diffusion term balances the
string depletion
effects
of the first term and the uniform density condition
for inflation (universe exponential expansion)
is satsfied. As the time elapses, however,
the depletion term in (\ref{denst})
dominates, the Universe's expansion is diminished gradually,
and eventually stops.
This is the case when the non-equilibrium non-critical string
approaches its (critical-string ) equilibrium state.
Hence, in our case one may
view (\ref{denst}) as an effective
model for the temporal evolution of the density of tubulin dimers.
Then, one can understand, at least qualitatively,
the above limitations of the MT growth process by the
presence of the kinks, since the latter can be associated with
an equilibrium ground state of the effective
string theory describing a MT.

\subsection{Other Experiments sensitive to Gravity}
\pr
The possibility that a (quantum) theory as weak as gravity affects
the physics of low-energy systems, like MT networks,
does not seem so remote, if we recall some recent
{\it experimental} indications \cite{tabony}
about an appreciable sensitivity
exhibited by MT structures to classical gravity effects,
and more general to weak external fields.
In such experiments, gravity effects can lead to
a sort of symmetry breaking and pattern formation
in assemblies of MT. Although theoretically
the sistuation is still very vague, however
the above phenomena appear consistent with
predictions based on reaction-diffucion
theories~\cite{react}, involving
out-of-equilibrium chemical reactions coupled to gravity.
As we have argued in ref. \cite{emn} quantum gravity in
non-critical string theory can be viewed as such
an out-of-equilibrium theory, resulting
in an irreversible flow of time and entropy production
at a fundamental (string) energy scale.
Whether MT systems,
whose (quantum) physics scale is that of electroweak effects,
are senstitive to {\it quantum} gravity effects,
as opposed to classical ones, still remains to be seen.
The idea does not seem so absurd
if we draw an analogy with what happens
in the neutral kaon system in particle physics~\cite{kaons}.
There, violations of quantum mechanics, associated with
quantum gravity effects of order $O[G_N^{\frac{1}{2}}m_K]
\simeq 10^{-19} $,
could be on the verge of being observed experimentally in $CPLEAR$
or $DA\phi NE$  facilities~\cite{kaons}.
\pr
In our picture, it becomes clear from the formal
disuccion in Appendix A (c.f.
(\ref{dollar},\ref{sel}) ),
that the effects that could lead to non-factorisability
of the target $S$-matrix, and therefore to quantum gravity
environmental entanglement, are suppressed by {\it a single}
power of $M_{String} \equiv (\sqrt{\alpha '})^{-1}$, and hence
the possibility of such terms having observable effects
may not be negligible.
\pr
This prompted us to identify above the scale $\alpha '$,
which appears as a typical scale of the MT system
represented as a string (\ref{ten}),
with the quantum gravity scale, $M_{Planck} \simeq 10^{19}~GeV$
(up to an uncertainty factor of ten, see discussion above).
This yields estimates of the collapse time
of order ${\cal O} (1 sec)$,
in brain MT neworks
consisting of $10^{12}$ MT dimers,
which is in excellent agreement with estimates of times
for conscious perception obtained by neurobiologists, based
on completely different methods. This offers support
to the above ideas, and to the ideas of ref. \cite{penrose,HP,dn},
that realistic quantum gravity effects may play an important
r\^ole in concious perception.
\pr
However, we should bear in mind that
in the case of systems pertaining to the
function of the brain things are by no means simple.
The simple fact that the collapse time, calculated on the basis
of string quantum gravity here, or conventional
quantum gravity~\cite{HP} (provided that the latter
exists as a mathematical theory), is in agreement with
estimates of conscious perception time obtained by quite
different methods, although a pleasant indication, however it
by no means constitutes a proof
of the relevance of quantum mechanics or quantum gravity
on brain function.
>From our point of view,
such a proof could come from observations of
fluctuations (`quantum jumps') of the length of isolated microtubules.
This is correlated strongly with the
stochastic growth or
saw-tooth behaviour of the length of a MT assembly, discussed
in the previous subsection.
When applied to an individual MT this approach may lead to
instabilities that could predict fluctuations of the tip
of MT due to quantum gravity entanglement~\cite{rosu}.
\pr
Whether experiments can be devised, which are
sensitive enough to capture such microscopic fluctuations
of isolated (cold) MT, is not known to us.
We believe, however, that if they could be devised, they
would constitute the best proof of the relevance of quantum
(gravity) effects for brain functioning, provided of course that
any other conventional source of mechanical instabilities~\cite{conv}
is excluded.
For instance,
the stochastic/diffusive nature
of Liouville gravity, advocated in ref. \cite{emn}, encourages
a comparison with the situation of ref. \cite{tabony} and,
in general, with experiments testing
predictions of reaction-difusion theories.
One is tempted to conjecture that the quantum
fluctuations of the tip of MT structures, predicted here
in the framework of Liouville theory, might also be seen in the
pattern formation of the experiments of ref. \cite{tabony}, provided
that the latter are repeated in `cold' environment so as to
minimize noise due to
(thermal) dissipation or other mechanical instabilities~\cite{conv},
that could interfere with pure
quantum gravity effects. Whether this is possible, or even conceivable
as an idea for future research, is unknown
to us at present, but we believe that
such speculations deserve closer
attention.
\pr
\section{Memory Coding and Capacity of the brain
as a (non-critical string) dissipative system }
\subsection{Existing local field theory models}
\pr
Irrespective of the possibility of proving experimentally
the possible effects of quantum gravity on brain function,
the conjecture of this work that
MT dynamics stems from one-dimensional
Ising spin chains in the brain, that can be represented
as a (completely integrable) non-critical string model
admitting space-time singularities,
implies certain peculiar but highly interesting
properties of brain functioning,
associated with non-equilibrium (dissipative) temporal evolution.
The latter implies an irreversible arrow ot time,
evolution of pure states into mixed ones, and, more generally,
what a particle field theorist would call
$CPT$ violation~\cite{emn}.
\pr
In this respect, our model has many things in common with
dissipative (local field theory) models of ref. \cite{umez,vitiello}
in an attempt to construct realistic models for memory capacity.
Below we shall
briefly review such models, and discuss the possible
advantages
of our (non-critical string)
approach over such local field theory approaches to brain function,
especially as far as
memory coding and capacity are concerned.
\pr
In conventional brain models~\cite{umez}, based on local field theories,
the kind of symmetry assumed is that of rotational electric dipole
symmetry of the surrounding water molecules.
The quantum numbers associated with the latter constitute
a certain class of code numbers.
If the brain lies on a specific ground state, which implies
spontaneous breaking of the dipole rotational symmetry, in order
to reach any other ground state corresponding to a new code number
it would require a sequence
of phase transitions that would destroy the previously stored
information, a procedure known
as {\it overprinting}~\cite{umez,vitiello}.
\pr
A way out of this problem of {\it memory capacity} would be to
increase the symmetry of the problem to the one with
huge dimensions~\cite{stewart}. In a
local field theory this cannot be done
without destroying the practical use of the model.
The problem is analogous to that of how to incorporate
the huge entropy of a macroscopic black hole in an information theory
framework within a local field theory setting. This again would
require an enormous amount of black hole degrees of freedom to
account for the macroscopic entropy, which would be hard, if not
impossible, to reconciliate with the finite number of degrees of
freedom existing in a local field theory.
\pr
An alternative approach,
is that of ref.
\cite{vitiello}, making use of
{\it dissipative} models for brain function, within a local field theory
framework. As observed in
ref. \cite{vitiello}, the doubling of degrees of freedom
which appears necessary for a canonical quantization
of an open system in a dissipative environment~\cite{umez},
is essential in yielding~\cite{harm} a {\it non-compact}
$SU(1,1)$ symmetry for the system of damped harmonic oscillators,
used as a toy example for simulating quantum brain physics.
The quantum numbers of such a system are the $SU(1,1)$ isospin
and its third component, $j \in Z_{\frac{1}{2}}, m \ge |j|$.
The memory (ground) state corresponds to $j=0$ and there is a huge
degeneracy  characterised by the various coexisting (infinite)
eigenestates
of the Casimir operator  for the $SU(1,1)$ isospin.
The open-character of the system introduces a time arrow
which is associated with the {\it memory printing} process and
is compatible with the `observation' that
`only the past can be recalled'~\cite{vitiello}.
As far as we can see,
the problem with this approach
is that it necessarily introduces
dissipation in the energy functional, through the non-hermitian
terms in the interaction hamiltonian between the subsystem and the
environment~\cite{umez,vitiello}.
Hence, it is not easy to see how to reconcile this
with the above-mentioned property of biological systems to
transfer energy without
dissipation across the cells~\cite{Frohlich,lal}.
Moreover, from our point of view, this approach cannot
take into account realistic quantum gravity effects,
which according to the hypothesis of the present work and
of refs. \cite{penrose,HP,dn}
are considered responsible for conscious perception.

\subsection{Advantages of a `stringy' representation of brain
models}
\pr
String theory seems to provide a way out of these problems~\cite{emn}
due to the infinite-dimensional
gauge stringy symmetries that mix the various levels.
In the (completely integrable)
black hole model of ref. \cite{witt}, which is used to simulate
the physics of the MT, there is an undelrying world-sheet
$SL(2,R)$
symmetry of the $\sigma$-model,
according to which the various stringy states are classified.
The various states of the model, including global string modes
characterised by discrete values of (target) energy and momenta,
are classified by the non-compact isospin $j$ and its
third component $m$, which - unlike the compact
isospin $SU(2)$ case - is not restricted by the  value of $j$.
Thus, for a given $j$, which in the case of string states plays the role
of energy, one can have an {\it infinity} of states labelled by the
value of the third isospin component $m$.
All such states are characterised by a $W_{1+\infty}$ symmetry
in target space. As we mentioned above, this symmetry
is responsible for the maintenance of
{\it quantum coherence} in the presence of a
black hole background~\cite{emn}, in the sense of an area-preserving
diffeomorphism in a matter phase-space of the two-dimensional target
space theory.
It should be noted that such area-preserving symmetries, as spectrum
generating algebras, also appear in connection with the
excitations of planar quantum Hall systems having non-degenerate
ground states~\cite{trugen}. So, it should not be considered
as a surprise that such symmetries appear in our two-dimensional
spin chain model for the brain MT. In the particular
case of two-dimensional string black holes there is even a
formal analogy with quantum Hall models,
as argued in ref. \cite{emnhall}.
\pr
In ref. \cite{emn} it has been argued that such symmetries
are responsible for an `infinite-dimensional' quantum hair ($W$-hair)
of the two-dimensional black hole, which consists of (conserved)
quantum (global) charges, like the $ADM$ mass~\cite{witt,adm},
characterising a black hole space-time
even asymptotically, i.e. after evaporation.
Such hair would induce a huge degeneracy in the ground state
of the system that could lead to the solution of the problem of
memory capacity. From a formal point of view, the rigorous
existence proof
of such conserved charges would be the
explicit construction of exactly marginal
deformations that correspond to turning on the above charges.
The exaclty marginal character of the deformations is
required in order to maintain
conformal invariance of the world-sheet $\sigma$-model and thus
stable ground state of the string.
\pr
As we mentioned in section 4.1,
for the black hole model of ref. \cite{witt},
used to simulate also the physics of the MT dynamics,
it is possible to construct~\cite{chaudh}
the exactly marginal deformation
corresponding
to the lowest non-trivial charge, which is the
$ADM$ mass of the black hole. From a $W_\infty$ symmetry
point of view, this would be the charge associated with the
spin-two part of the target space spectrum, i.e. the stress-energy
tensor of the black hole.
There is a huge degeneracy of the ground state of the system
which is due to the existence of known exactly marginal deformations
that are responsible for changing continuously the
$ADM$ mass of the black hole, as discussed in section 4.1.
In the notation of ref. \cite{chaudh}, such deformations
are denoted by $L_0^2 {\overline L}_0^2$, as we mentioned previously.
Their coupling constant,
which
is a {\it free} parameter of the string model, shifts
the $ADM$ mass of the black hole space time.  The above operator
turns on only backgrounds corresponding to the (discrete)
higher-level string states that do not propagate in space-time.
The ground state of such models consists of turning on
backgrounds corresponding to matter propagating states.
Such backgrounds are turned on by the other known
exactly marginal
deformation $L_0^1 {\overline L}_0^1$, which
mixes
the propagating states (belonging to the lowest string level)
with an infinity of higher-level string global states.
Both operators owe their existence to
the target space $W_{1+\infty}$-spectrum-generating algebra
of the black hole space-time~\cite{emn,chaudh}. The latter
is broken explicitly by
`measurement' by
local scattering experiments or in general by operations
that are performed within localised regions of space-time,
such as those taking place in the conscious part of the brain.
\pr
Such a procedure will integrate out the global degrees of freedom,
leaving only an effective (string) theory of propagating degrees
of freedom in a black hole background space time.
For each matter ground state of a propagating degree of freedom,
say the zero mode of the massless field corresponding to the static
``tachyon'' background of ref. \cite{witt}, with $SL(2,R)$
quantum numbers
$j=-\frac{1}{2}, m=0$, there will be an infinite degeneracy
corresponding to a continuum of black hole space-time backgrounds
with different $ADM$ masses.
These backgrounds are
essentially generated by
adding various constants to the configuration of the dilaton
field in this
two-dimensional string theory~\cite{witt,chaudh}.
It should be noted here that the infinity of
propagating
``tachyon'' states (lowest string mass-level (massless) states),
corresponding to other values of $m$, for
continuous representations of $j$, constitute
{\it excitations} about the ground state(s), and, thus,
they should not be considered as contributing to the
ground state degeneracy.
In principle, there may be an additional infinity of
quantum numbers corresponding to higher-level $W$-hair charges
of the black hole space time which are
believed~\cite{emn}
responsible for quantum coherence at the full string theory level.

\subsection{Memory Coding and capacity of the `stringy' MT model}
\pr
Taking into account the conjecture of the present work,
that formation of virtual black holes
can occur in brain MT models, which
would correspond to different modes of collapse of
pulses of the displacement field $\psi $
defined in (\ref{eight}),
one obtains a system of {\it coding} that is capable to
solve in principle the problem of memory capacity.
Information is stored in the brain in the following sense:
every time there is an external stimulus that brings the brain
out of equilibrium, one can imagine an abrupt conformational
change of the MT dimers, leading to a collapse
of the pulse pertaining to the displacement field. Then a (virtual)
black hole
is formed leading to a spontaneous collapse of the MT network
to a ground state characterised by say a special configuration
of the displacement field $(j, m)$. This ground state
will be conformally invariant, and therefore a true vacuum
of the string, only after complete evaporation of the black hole,
which however would keep  memory of the particular collapse
mode in the `value' of the constant added to the dilaton field,
or other $W$-charges.
This
reflects the existence of additional exactly marginal
deformations, consisting of global modes only,
that are not direclty accessible by local scattering experiments,
in the context of the low energy theory of propagating modes
(displacement field configurations $\psi $).
In such a case, the resulting ground state will be infinitely
degenerate, which would solve the problem of {\it memory
capacity}\footnote{It should be noted that the
picture of the formation and evaporation of the two-dimensional
black holes (foam)~\cite{emn},
that we advocate in this work in connection
with concious brain processes, is not unrelated
to the recently developed approach to space-time
foam in the context of four-dimensional quantum
string theory~\cite{strom}, employing duality
symmetries. The latter help in making exactly
solvable
a
strongly-coupled problem, like the formation and
evaporation of a (virtual) space-time singularity in string theory,
by mapping it to a weakly-coupled
$D$-brane theory.
In this picture, virtual black holes in
string loops are believed responsible
for the transition among string vacua, a process
which in our two-dimensional case
case would correspond to the
evaporation of a black hole via world-sheet
instanton effects, as discussed in section 4.
However, we should stress that our approach
employs time {\it dynamically} through the
{\it non-criticality} of the strings involved,
something which had not been considered so far
in the approach of ref. \cite{strom}.}.
\pr
Breaking of this degeneracy, can be achieved
by means of an
{\it external stimulus}, which is believed to be
due to a weak field~\cite{delgiud}. For instance,
following the suggestion of ref. \cite{delgiud}, we
may imagine that an external weak field produces a
spontaneous breaking of the electric dipole rotational
symmetry in the water molecules, resulting in a
`lasering'~\cite{prep}
of the environmental surroundings of the MT system
(excitation of coherent dipole quanta).
Such an excitation of collective modes results
in a specific {\it code} characterising the ground state,
as we mentioned above.
The so selected
ground state of the ordered water molecules affects the
MT chains, due to the friction coupling $\rho $ in (\ref{ten}),
(\ref{dilaton}). As becomes clear from the analysis
of section 2 ((\ref{ten})-(\ref{13})),
the effects of the environment are described by selecting
a specific value for the vacuum expectation value (condensate)
of the dilaton field in our suggested stringy approach to MT
dynamics.
The other $W$-charges (moduli)
may also be selected this way,
which we believe corresponds to the {\it memory printing}
process, i.e. storage of information by a selection
of a given ground state. A new information would then
choose a different value of the dilaton field
or other $W$-hair charges,
etc. This provides a new and satisfactory mechanism of {\it memory
recall} in the following sense: if a new pulse happens to
correspond to the same set of (conserved)
$W$-hair moduli
configurations~\cite{emn}, then the associated virtual black hole
will be characterised by the same set of quantum hair, and
then the same memory state is reached {\it asymptotically}
(process of `memory
recall'). The discussion we gave in the previous section
about the r\^ole of
world-sheet instanton deformations in shifitng the
$ADM$ mass of a black hole
(\ref{adm}), while keeping memory of the dilaton v.e.v.,
finds a natural application in this coding
process. Moreover,
the irreversible arrow of time, endemic in Liouville
string theory~\cite{emn} explains naturally why ``only
the past can be recalled''~\cite{vitiello}.
\pr
To understand why the above process leads to a special coding,
and how time reversal is spontaneously broken, as a result
of this coding, it is sufficient to recall our
discussion
above, according to which in
the presence of a space-time foamy environment,
characterised by the virtual appearence and evaporation of black holes,
there is a coupling of global modes to the propagating modes.
As a result of the exactly marginal character of the
deformations~\cite{emn,chaudh},
which thus respect conformal invariance at a string level,
the environmental global
modes match in a special way with the propagating mode  $j=-\frac{1}{2}$
$m=0$, which is the
zero mode of the (massles)
tachyon corresponding to the tachyon background
of a two dimensional black hole which constitutes the {\it ground}
state or {\it memory} state of our system.
This is a {\it special
coding} which were it not for the infinte degeneracy
of the black hole space time would lead to
a restricted memory capacity\footnote{It should
be noted, once more, that the various other
(infinite) states corresponding to continuous representations
of the $SL(2,R)$ symmetry that pertain to various
tachyon modes do not constitute memory states,
because, as mentioned earlier,
they are just {\it excitations} about the ground state.}.
\pr
We cannot resist in
pointing out that the
existence of such coded situations in memory cells
bears an interesting resemblence with DNA coding,
with the important difference, however, that
here it occurs in the model's state space.
In this context, we note that the genetic diversity
is not due to an infinite number of nucleotide
types, since in nature there are only four of them,
paired by two ($A~=~T/G~\equiv~C$), but
rather to a macroscopically
large number of existing combinations in the DNA helix.
Similarly, for the extremely rich (macroscopic) memory
capacity
in our stringy MT model, we may not need
the full
(infinite) set of the $W$-hair charges,
but just the
dilaton vev $<\Phi >$ may be sufficient as a `collective'
mode. The latter is, as we mentioned earlier, related
to external stimuli through the equations (\ref{ten})-(\ref{13}).
\pr
Before closing, we consider it as useful to compare once more
our results
with those of ref. \cite{vitiello}, using a dissipative
system approach to the brain memory problem, but within a local
field theory framework.
In both models, the effect of the `environment' was crucial
in providing an {\it infinity} of degrees of freedom, capable of
solving the memory capacity problem.
The crucial difference of our
string case is that the ground state of the string system, which
is conformally invariant, is actually a state with given quantum
numbers $j=-\frac{1}{2}, m=0$ of
the $SL(2,R)$ isospin in the asymptotically
flat space time case.
The degeneracy occurs, as we have already mentioned, as a result
of the `existence of an environment' of global modes, inaccessible
by local scattering operations of the brain, which lead to
exactly marginal deformations shifting the ground state
value of the dilaton field, or in general leading
to an infinity of $W$-hair charges.
Moreover, the effects of the environment in the string
case, as contrasted to that of ref. \cite{vitiello},
are such
that energy is {\it conserved on the average}, thereby making
the connection with energy-loss-free energy transport
across biological structures more apparent.
The importance of the
coupling between global and localized (propagating)
modes of this (1+1)-dimensional string theory
lies on the fact that it
leads to a time arrow for {\it specifically stringy
reasons}~\cite{emn}.
Essentially, time arrow emerges as a result of
{\it loss of information}, carried by the global modes
which are inaccessible to localized scattering processes,
such as the ones taking place in
the brain. From a target space-time point of view, such an information
loss is manifested through entropy production and
non factorisability (c.f.
eq. (\ref{dollar}) of Appendix A)
of the superscattering
operator~\cite{emn}.
\pr
\pr
\section{Conclusions}
\pr
In this work we have argued, following
general ideas in
refs. \cite{penrose,HP},
that
in certain parts
of the brain
there is the possibility of the formation
of quantum superpositions
which extend over reasonably
large distances and time scales. Such superpositions
imply probably
that brains cannot be understood as classical computers.
We have made attempts to locate a possible microscopic
mechanism
for the formation of such superpositions, and argued
that a natural place for their appearance are the networks
of Microtubules (MT) that exist in the brain.
Moreover we have considered in some detail the
r\^ole of quantum gravity environmental
effects in
destroying such quantum-coherent superpositions
(collapse of the wave function), thereby
leading to conscious perception.
\pr
We have presented an effective model for the
simulation of the dynamics of the tubulin dimers
in the brain. We have used an effective $(1 + 1)$-dimensional
string representation to
study the dynamics of
a detailed mechanism
for energy transfer in the
biological cells. We argued how it can give rise to
a large-scale coherent state in the dimer lattice.
Such a state
is obtained from quantization of
kink solitonic states that transfer energy
through the cell without dissipation.
The quantization became possible
through the freedom that string theory offers, enabling
one to
cast dynamical problems with friction in a
Hamiltonian form.
\pr
The collapse phenomenon in our approach does
not require the existence of a wave-fucntion,
and it
is induced by the formation of microscopic
black holes (singularities)
in the effective one-dimensional
space-time of the tubulin chains. This is achieved
by the dynamical collapse of pulses of the
displacenent field of the MT dimers. The pulses are a result of
abrupt conformational changes ($\alpha\leftrightarrow\beta$)
that sufficiently
distort the surrounding space-time.
In this sense, the situation is similar but not identical, to
the `sudden hits' that a particle's wave function suffers
occasionally (fixed `by hand' to occur
every $10^8$ years)
in the model
of quantum measurement
of Girardi, Rimini and Weber \cite{grw}.
However, contrary to these conventional theories,
our stringy approach to gravity-induced collapse
\cite{emn}
incorporates automatically
an irreversible flow of time
for specifically {\it stringy} reasons, and {\it
energy conservation}, while it provides
a dynamical determination of the decoherence
or collapse time, depending on Newton's constant
$G_N$ and the `energy' content of the system.
\pr
When applied to the model of MT, our approach
implies a collapse time of
${\cal O}(1\,{\rm sec})$,
which is obtained by the interaction of a tubulin dimer with
a fraction of $10^{-7}$
of the total number of tubulin dimers in the brain.
This number is fairly close to the fraction of the
brain that neuroscientists believe responsible
for human perception.
This is a very strong indication
that the above ideas, although speculative at this stage,
might be relevant for the discovery
of a physical model for consciousness and its relation
to the irreversible flow of time.
In addition, our model predicts
damped
(microscopic)
quantum-gravity-induced oscillations
of the length of isolated MTs, which
are due to the different properties
of the two tubulin conformations
under polymerization (phenomenon
of bounded dynamical-instability growth).
\pr
There are certain formal aspects of our effective model,
namely its two-dimensional structure and its complete
(quantum) integrability, that might turn out to be
important
features for the construction of realistic soluble
models for
brain function. The quantum
integrability is due to generalized infinite-dimensional
symmetry structures ($W$-symmetries and its generalizations)
which are strongly linked with issues
of quantum coherence and unitary evolution in phase-space.
Moreover we have seen that due to such huge stringy symmetries
there is an automatic solution of the memory coding and
capacity problems, which some local field theories models
of the brain
are plagued with.
Such symmetries are related to global non-propagating modes
of the effective string theory, which do not decouple
from the propagating (observed) modes in the presence of
(microscopic)
space-time singularities.
It  will be interesting to
understand further the physical r\^ole of such structures
in the models of MT. At present they appear as an environment
of fundamental string modes that are physical at Planck scales.
However, such structures, may admit a less-ambitious
physical meaning, associated with fundamental biological
structures
of the
brain.
We should stress that
the entire picture of non-critical
string we have described above, which is a
model-independent picture as far as environmental
operators are concerned, could still apply in such cases,
but
simply describing
purely biological
environmental entanglement of the conscious part
of the brain, the latter still being described as a
completely integrable model.
\pr
In this respect,
a possible connection of the brain MT system
with conformal models exhibiting
{\it disorder} has also been anticipated, which might
be a key feature for a rigorous formulation
of the concious perception process from a conformal
field theory point of view.
This has been associated with the fact that
the formation and evaporation of the microscopic
black holes in the MT chains, which are the result
of external stimuli, induce disorder due to back-reaction/recoil
effects of
matter (displacement field of MT)
on the effective two-dimensional space time of the chains,
expressing environmental entanglement.
This picture is in general agreement with the
recently developed picture of space time foam
in four-dimensional string theory, employing
duality symmetries to map the strongly-coupled
system of
(virtual) black hole
singularities
to well-behaved weakly coupled $D$-brane theories~\cite{strom}.
\pr
We believe that
our work, although admitedly speculative at this stage,
motivates further studies
along the above directions which might prove useful
towards a possible understanding of physical processes
governing brain functions. In our opinion, the most realistic
and profitable
approach one could follow
would be that of lattice simulations of such systems.
This will allow for a rigorous study of {\it finite size}
effects, which have been neglected in the present
continuous formalism. Such effects may play an important
r\^ole in determining the {\it limitations}
of the systems as quantum computers, as
far as properies such as:
memory capacity, coding,
and even the formation
of MT structures themselves,
are
concerned. We hope to come back to such issues
in the future.
\pr
\nk {\Large{\bf  Acknowledgements}}
\pr
It is a pleasure to acknowledge discussions
with J. Ellis.
We also acknowledge useful discussions and
correspondence with
S. Hameroff. N.E.M. acknowledges informative communications
with H. Rosu on quantum jumps and experimental observations of
dynamical
instabilities in nanotubes.
The work of N.E.M. is supported by a EC Research Fellowship,
Proposal Nr. ERB4001GT922259 .
That of D.V. N. is partially supported by D.O.E. Grant
DEFG05-91-GR-40633.
\pr
\newpage

\nk {\Large {\bf Appendix A} }
\pr
\nk {\bf Extracts from
Non-Critical (Liouville) String Theory  and  Time
as the Liouville scale}
\pr
In this appendix we comment on the non-factorizability
of the induced target-space $\nd{S}$-matrix for matter
scattering in a non-conformal string background.
This reflects information leakage as a result of the
non-critical character of the string. Although  for our purposes,
primarily, we shall be interested in a specific string background,
that of a stringy black hole, however in this section
our discussion will be kept as general as possible with the
aim of demonstrating the generality of our scheme.
\pr
Consider a conformal field theory on a two-dimensional
world sheet, described by an action $S[g^*]$.
The $\{ g^* \}$  are a set of space-time backgrounds.
The theory is perturbed by a deformation $V_i$,
which is not conformal invariant
\be
   S[g] = S[g^*] + \int d^2 z g^i V_i
\label{C1}
\ee
The couplings $g^i$ corrspond to world-sheet
renormalization group $\beta$-functions
\be
   \beta ^i = (h_i - 2)
   (g^i - (g^{*})^i) +
c^i_{jk}(g^j - (g^{*})^j) (g^k - (g^{*})^k) +  \dots
\label{C2}
\ee
expressing the scale dependence of the non-conformal
deformations.
The operator product expansion coefficients
are defined as usual  by coincident limits in the product
of two vertex operators $V_i$
\be
lim_{\sigma \rightarrow 0} V_j (\sigma )V_k (0) \simeq
c^i_{jk} V_i (\frac{\sigma}{2}) + \dots
\label{C3}
\ee
where the completeness of the set $\{ V_i \}$ is assumed.
\pr
Coupling the theory (\ref{C1}) to two-dimensional quantum gravity
restores the conformal invariance at a quantum level,
by making the gravitationally-dressed operators
$[V_i]_{\phi}$ {\it exactly} marginal, i.e. ensuring
the absence of any covariant
scale dependence with respect to the
world-sheet metric $\gamma _{\alpha\beta}$.
Below we simply outline the basic results,
used in our approach here.

One rescales the world-sheet metric
\be
    \gamma _{\alpha\beta} = e^{\phi} {\widehat \gamma}_{\alpha\beta}
\label{C4}
\ee
with ${\widehat \gamma}$ is kept fixed,
and then one integrates over the Liouville mode $\phi $.
The measure of such an integration\cite{DDK}
can be expressed in terms of the fiducial metric
${\widehat \gamma }$ by means of a determinant
which is the exponential of the Liouville action.
The final result for the gravitationally-dressed
matter theory is then
\be
S_{L-m} = S[g^*] + \frac{1}{4\pi\alpha '}
\int d^2 z \{\partial _\alpha \phi \partial ^\alpha \phi
- QR^{(2)} + \lambda ^i(\phi ) V_i \}
\label{C5}
\ee
where $\alpha '$ is the Regge slope for the world-sheet theory
(inverse of the string tension).
The gravitational dressing of the operators follows
from the requirement of restoring
the conformal invariance of the theory, at
any given order in the coupling-constant
expansion. For instance, to order $O(g^2)$
the gravitationally-dressed
coupling $\lambda (\phi )$ are given by\cite{DDK,schmid}:
\be
\lambda ^i(\phi ) =g^i e^{\alpha _i \phi }
+ \frac{\pi}{Q \pm 2\alpha _i } c^i_{jk} g^jg^k
\phi e^{\alpha _i \phi } + \dots
\label{C6}
\ee
with
\be
Q=\sqrt{\frac{|25-c|}{3}}  \qquad ; \qquad
\alpha _i ^2 + \alpha _i Q =sqn(25-c)(h_i - 2)
\label{C7}
\ee
and $c$ is the (constant) central charge of the
non-critical string.
>From the quadratic equation for $\alpha _i$
only the solution
\be
\alpha _i = -\frac{Q}{2} +
\sqrt{\frac{Q^2}{4} - (h_i - 2)}
\label{solutions}
\ee
for $c \ge 25 $, is kept due to
the Liouville
boundary conditions.

In ref. \cite{emn} we made an extra assumption, as compared
to the above standard Liouville dynamics. We identified
the field $\phi $ with a dynamical local scale on the
world sheet. This induces extra counterterms
in the world-sheet renormalized action. Consistency
of the scheme required that the Liouville $\beta $
functions are identical with the flat space renormalization
coefficients upon the replacement $g^i \rightarrow \lambda (\phi)^i$.

The type of operators that we are interested in this
work, are such that $h_i =2$ but $c^i_{jk} \ne 0$.
In the language of conformal field theory
this means that these operators are $(1, 1)$
but {\it not exactly marginal}.
>From (\ref{C6}), then, one obtains the simple relation
\be
   \frac{d \lambda ^i(\phi) }{d t_p} =  \beta ^i
\label{C8}
\ee
where the time $t_p$ is related to the Liouville mode $\phi $
as
\be
        t_p = -\frac{1}{\alpha Q}{\rm ln}A \qquad ; \qquad
  A \equiv \int d^2z \sqrt{{\hat \gamma}}
e^{\alpha \phi (z,{\bar z})}
\qquad ; \qquad \alpha = -\frac{Q}{2} + \frac{1}{2}\sqrt{Q^2 + 8}
\label{tphys}
\ee
with $A$ the world-sheet area.
In the local scale formalism of ref. \cite{emn}
$Q$ is given by
\be
Q=\sqrt{\frac{|25-C[g,\phi]|}{3} + \frac{1}{2}\beta ^i G_{ij} \beta ^j}
\label{C9}
\ee
where $C[g,\phi ]$ is the Zamolodchikov $C$-function
\cite{zam}, which reduces to the
central charge $c$ at a fixed point of the flow.
The extra terms in (\ref{C9}), as compared to (\ref{C7}), are due to
the local character of the renormalization group scale\cite{emn}.
Such terms may always be removed by non-standard redefinitions
of $C[g,\phi ]$. The quantity
$G_{ij}$ is related to divergencies
of the two-point functions $<V_iV_j>$
and hence to Zamolodchikov metric
\cite{zam,emn}.

>From the renormalization-group structure
(\ref{C6}) one obtains close to a fixed point \cite{tseytl}
\be
  {\ddot \lambda (\phi )}^i + Q {\dot \lambda} ^i = -\beta ^i = - G^{ij}
  \partial _i C [\lambda, \phi]
\label{C10}
\ee
where the dot denotes differentiation
with repsect to the Liouville local scale $\phi $.

For the $C[g,\phi ]$  (local in target space-time)
one obtains near a fixed point
\be
 {\ddot C}[g, t]
 + Q [g,t] {\dot C} [g,t] \le 0 ~for C \ge 25
 \qquad ; \qquad  Q^2 [g, t] =\frac{1}{3}
(C[g,t ] - 25)
\label{C11}
\ee
The small oscillations of $C[\lambda, \phi ]$,
before it settels down to a fixed point,
are due to the `non-unitary' world-sheet contributions
of the Liouville mode $\phi$ ; however
globally in target space-time
there is a monotonic change of the degrees of freedom of the
system, as discussed in detail in \cite{emn}.

These considerations can be understood more easily
if one looks at the correlation functions
in the Liouviulle theory, viewing the Liouville field
as a local scale on the world sheet .
Standard computations\cite{goulian} yield for an $N$-point correlation
function among world-sheet integrated
vertex operators $V_i\equiv \int d^2z V_i (z,{\bar z}) $ :
\be
A_N \equiv <V_{i_1} \dots V_{i_N} >_\mu = \Gamma (-s) \mu ^s
<(\int d^2z \sqrt{{\hat \gamma }}e^{\alpha \phi })^s {\tilde
V}_{i_1} \dots {\tilde V}_{i_N} >_{\mu =0}
\label{C12}
\ee
where the tilde denotes removal of the
Liouville  field $\phi $ zero mode, which has been
path-integrated out in (\ref{C12}).
The world-sheet scale $\mu$ is associated with cosmological
constant terms on the world sheet, which are characteristic
of the Liouville theory \cite{DDK}.
The quantity $s$ is the sum of the Liouville anomalous dimensions
of the operators $V_i$
\be
s=-\sum _{i=1}^{N} \frac{\alpha _i}{\alpha } - \frac{Q}{\alpha}
\qquad ; \qquad \alpha = -\frac{Q}{2} + \frac{1}{2}\sqrt{Q^2 + 8}
\label{C13}
\ee
The $\Gamma $ function can be regularized\cite{kogan,emn}
(for negative-integer
values of its argument) by
analytic coninuation to the complex-area plane using the
the Saaschultz contour
of Fig. 4. This yields the possibility
of an increase of the running central charge
due to the induced oscillations of the dynamical
world sheet area (related to the Liouville zero mode).
This is associated with the oscillatory solution
(\ref{C11}) for the Liouville central charge.
On the other hand, the bounce intepretation
of the infrared fixed points of the flow,
given in refs. \cite{kogan,emn},
provides an alternative picture
of the overall monotonic change
at a global level in target space-time.

The above formalism also allows for an
explicit demonstration of the
non-factorizability of the
superscattering matrix associated with
target-space interactions in non-critical
string theory. This
was very important for our purposes
in the context of the collapse
of the wave-function as a result of quantum
entanglement due to quantum gravity fluctuations.

To this end, one expands the Liouville
field in (normalized) eigenfunctions  $\{ \phi _n \}$
of the Laplacian $\Delta $ on the world sheet
\be
 \phi (z, {\bar z}) = \sum _{n} c_n \phi _n  = c_0 \phi _0
 + \sum _{n \ne 0} \phi _n \qquad \phi _0 \propto A^{-\frac{1}{2}}
\label{C14}
\ee
with $A$ the world-sheet area,
and
\be
   \Delta \phi _n = -\epsilon_n \phi _n  \qquad n=0, 1,2, \dots,
\qquad \epsilon _0 =0
\qquad (\phi _n, \phi _m ) = \delta _{nm}
\label{C15}
\ee
The result for the correlation functions (without the Liouville
zero mode) appearing on the right-hand-side of eq. (\ref{C12})
is, then
\bea
{\tilde A}_N \propto &\int & \Pi _{n\ne0}dc_n exp(-\frac{1}{8\pi}
\sum _{n\ne 0} \epsilon _n c_n^2 - \frac{Q}{8\pi}
\sum _{n\ne 0} R_n c_n + \nn \\
~&~&\sum _{n\ne 0}\alpha _i \phi _n (z_i) c_n )(\int d^2\xi
\sqrt{{\hat \gamma }}e^{\alpha\sum _{n\ne 0}\phi _n c_n } )^s
\label{C16}
\eea
with $R_n = \int d^2\xi R^{(2)}(\xi )\phi _n $. We can compute
(\ref{C16}) if we analytically continue \cite{goulian}
$s$ to a positive integer $s \rightarrow n \in {\bf Z}^{+} $.
Denoting
\be
f(x,y) \equiv  \sum _{n,m~\ne 0} \frac{\phi _n (x) \phi _m (y)}
{\epsilon _n}
\label{fxy}
\ee
one observes that, as a result
of the lack of the zero mode,
\be
   \Delta f (x,y) = -4\pi \delta ^{(2)} (x,y) - \frac{1}{A}
\label{C17}
\ee
We may choose
the gauge condition  $\int d^2 \xi \sqrt{{\hat \gamma}}
{\tilde \phi }=0 $. This determines the conformal
properties of the function $f$ as well as its
`renormalized' local limit\cite{lag}
\be
   f_R (x,x)=lim_{x\rightarrow y } (f(x,y) + {\rm ln}d^2(x,y))
\label{C18}
\ee
where  $d^2(x,y)$ is the geodesic distance on the world sheet.
Integrating over $c_n$ one obtains
\bea
~&& {\tilde A}_{n + N} \propto
exp[\frac{1}{2} \sum _{i,j} \alpha _i \alpha _j
f(z_i,z_j) + \nn  \\
~&&\frac{Q^2}{128\pi^2}
\int \int  R(x)R(y)f(x,y) - \sum _{i} \frac{Q}{8\pi}
\alpha _i \int \sqrt{{\hat \gamma}} R(x) f(x,z_i) ]
\label{C19}
\eea

We now consider
infinitesimal Weyl shifts of the world-sheet metric,
$\gamma (x,y) \rightarrow \gamma (x,y) ( 1 - \sigma (x, y))$,
with $x,y$ denoting world-sheet coordinates.
Under these,
the correlator $A_N$
transforms as follows\cite{emnest}
\bea
&~&
\delta {\tilde A}_N \propto
[\sum _i h_i \sigma (z_i ) + \frac{Q^2}{16 \pi }
\int d^2x \sqrt{{\hat \gamma }} {\hat R} \sigma (x) +    \nn \\
&~&
\frac{1}{{\hat A}} \{
Qs \int d^2x \sqrt{{\hat \gamma }} \sigma (x)
       +
(s)^2 \int d^2x \sqrt{{\hat \gamma }} \sigma (x) {\hat f}_R (x,x)
+  \nn \\
&~&
Qs \int \int d^2x d^2y
\sqrt{{\hat \gamma }} R (x) \sigma (y) {\hat {\cal
 G}} (x,y) -
  s \sum _i \alpha _i
  \int d^2x
  \sqrt{{\hat \gamma }} \sigma (x) {\hat {\cal
 G}} (x, z_i) -   \nn \\
&~&
 \frac{1}{2} s \sum _i \alpha _i{\hat f}_R (z_i, z_i )
  \int d^2x \sqrt{{\hat \gamma }} \sigma (x)
-    \nn \\
&~&
 \frac{Qs}{16\pi} \int
  \int d^2x d^2y \sqrt{{\hat \gamma (x)}{\hat \gamma }(y)}
  {\hat R}(x) {\hat f}_R (x,x) \sigma (y)\} ] {\tilde A }_N
\label{dollar}
\eea
where the hat notation denotes transformed quantities,
and
the function  ${\cal G}$(x,y)
is defined as
\be
  {\cal G}(z,\omega ) \equiv
f(z,\omega ) -\frac{1}{2} (f_R (z,z) + f_R (\omega, \omega ) )
\label{C20}
\ee
and transforms simply under Weyl shifts\cite{lag}.
We observe from (\ref{dollar}) that
if the sum of the anomalous dimensions
$s \ne 0$ (`off-shell' effect of
non-critical strings), then there are
non-covariant terms in
(\ref{dollar}), inversely proportional to the
finite-size world-sheet area $A$.
In general, this is a feature of non-critical strings
wherever the Liouville mode is viewed as a local scale
of the world sheet. In such a case,
the central charge of the theory
flows continuously with time/scale $t$,
as a result of the Zamolodchikov
$c$-theorem \cite{zam}. In contrast, the screening operators
yield quantized values\cite{aben}.
This induced  time ($A$-) dependence
of the correlation function $A_N$ implies the
breakdown of their interpretation as
factorisable $\nd{S}$-matrix elements.
\pr
In our framework, the effects of the quantum-gravity
entanglement induce such $A$-dependences
in correlation functions of the propagating
matter vertex operators of the string \cite{emn},
correspondng to the displacement field $u(x,t)$ of the MTs.
To this end, we first note that
the physical states in such completely integrable
models fall into representations
of the $SL(2,R)$ target symmetry, which are classified by
the non-compact isospin $j$ and its third component
$m$\cite{distler}. There is a formal
equivalence of the physical states
between the flat-space time $(1 + 1)$-dimensional string
and the black-hole model\footnote{Originally, there were claims
\cite{distler}
that there are extra states
in the black hole models, as compared to the flat-space time
string; however, later on it has been shown that
such states
can be either
gauged away~\cite{eguchi} or
boosted~\cite{ard},
and so they disappear from the physical spectrum.},
which confirms the point of view\cite{emnsel,emn}
that the flat-space $(1+1)$-dimensional string
theory is the spatially- (and temporally-) asymptotic limit of the
$SL(2,R)/U(1)$ black hole.
The existence of discrete (quasi-topological, non-propagating)
Planckian modes in the two-dimensional string theory
leads to {\it selection} rules\cite{emnsel}
in the number $N$ of the scattered propagating degrees of freedom,
according to the intermediate-exchange state :
\bea
~&~&{\rm Asymptotic~correspondence}~(\epsilon _\phi~(p) \equiv {\rm
Liouville~energy~(momentum)}): \nn \\
~&~&j \rightarrow \epsilon_\phi ,~m \rightarrow
  \frac{3 p}{2\sqrt{2}}
\nn \\
~&~&{\rm Asymptotic~Kinematics}:
\nn \\
~&~&\sum _{i=1}^{N-1} p_i=\frac{N-2}{\sqrt{2}} \qquad ; \qquad
p_N =-\frac{N-2}{\sqrt{2}} \qquad; \qquad N \ge 3
\nn \\
~&~&{\rm Selection~Rules}: \nn \\
~&~&j=\frac{1}{3}m - 1 + \frac{1}{2} (N -2)~,~j \ge \frac{1}{4}(N- 5)
\label{sel}
\eea
Such rules are obtained by imposing the Liouville
energy $\epsilon _\phi $ and momentum $p$
conservation, leading to $s =0$, with $s$ the sum of
Liouville anomalous dimensions as defined earlier.
Obviously, if the exchange state is an (off-shell)
propagating
mode, belonging to the continuous representation
of $SL(2,R)$, i.e. $j \in {\bf R}$, $j \ge -\frac{1}{2}$, there
are no restrictions
on $N$, and a convetnional $S$-matrix amplitude
can be defined as the residue of the Liouville
amplitudes with respect to the single poles in $s$~\cite{legpoles}.
However, in two space-time dimensions graviton excitations
are {\it discrete}, corresponding to string-level-one
representations  of $SL(2,R)$.
Hence,  once non-trivial quantum-gravity  fluctuations
are considered in our approach, which in two dimensions
are black-hole backgrounds (\ref{st11}),
one has to take into account  discrete
on-shell exchange modes in
the Liouville correlation functions\footnote{The on-shell condition
imposes algebraic relations among $j$ and $m$ for such modes,
involving the string-level number due to the Virasoro constraints
\cite{distler}. This, in turn, implies restrictions
to the number $N$
of scattered particles in such cases.}. Such states represent
{\it excited} states of the (virtual) black-holes, created
by the collapse of the propagating matter modes $u(x,t)$,
as described in section 4 (\ref{st11}).  In two-dimensional
string theory black holes are like particles\cite{emn}, the
difference being their topological nature.
Such modes constitute, in our case, `the consciousness
degrees of freedom', which cannot be measured
by local scattering experiments. Integrating them out
in the `mind', implies a time arrow as described in ref.
\cite{emn}. Indeed, in the correlation functions
(\ref{C12}),
as a result of Liouville energy conservation
(\ref{sel}), one of the modes is necessarily discrete.
If we suppress such modes, and consider only
external propagating modes, accessible to
physical scattering processes,
then it is evident
that $s \ne 0$. According to our previous
analysis (\ref{dollar}), then, this
implies world-sheet-area($A$) dependence
of the correlation functions.
In this picture,
we also note that
Quantum-Gravitational fluctuations of singular space-time
form, correspoding to higher-genus world-sheet effects,
have been argued~\cite{emn} to be represented
{\it collectively}
by world-sheet instanton-anti-instanton deformations
in the stringy $\sigma$-model.
It is known \cite{yung} that such configurations
are responsible for a {\it non-perturbative} breakdown
of the conformal invariance of the $\sigma$-model.
Using a dynamical (world-sheet) renormalization-group
scale (Liouville mode) to represent
all
such non-conformal invariant
effects\cite{emn}, and identifying it
with the target time,
one, then, arrives
at non-factorisable superscattering
operators, as described above.
\pr
Notice that the precise microscopic
nature of the environmental operators is not
essential as long as the latter imply a conformal
anomaly. There are general consequences of this
conformal anomaly,
including dynamical collapse of the string theory
space to a certain configuration, as discused in
section 4. A similar situation occurs in
ordinary
quantum mechanics of open systems.
Once a stochastic framework
using state vectors
is adopted \cite{gisin} for the
description of environmental effects,
there will always be localization
of the state vector in one of its
channels, irrespective of the detailed form of the
environment operators.
It should be noted that stochasticity
is a crucial feature of our approach
too.
This follows from the stochastic nature of
the renormalization group in two-dimensions
\cite{friedan,emn}.
This stochasticity was argued to play an important r\^ole
on the transition from quantum to classical worlds
in the brain (conscious perception), discussed  in section 4, as
well as
in the MT growth, discussed in section 5.
\pr
\nk {\bf World-sheet
Instanton Calculus and evaporation of a two-dimensional
Stringy Black hole}
\pr
In this part of the appendix we would like
discuss briefly some technical but important
apsects of the specific two-dimensional
black hole model of ref. \cite{witt}, which
is the exact conformal field theory
used in the simulation of the dynamics of the
MT chains in this work. In particular, we shall
outline some details pertaining to
world-sheet instantons effects
in yielding
{\it extra logarithmic divergencies} in
correlation functions of discrete matter
operators (`tachyons') of the two-dimensional
theory. This justifies our assumption in the
text that the effects of instantons are associated
with topological (zero) modes of the black hole model,
whose circulation along thin tudes of long handles
connecting Rieman surfaces produces extra
divergencies, expressed by the addition of
$\theta$-terms in the effective action.
We shall be very brief below, and concentrate only
on giving the basic results. For details on the derivation
we refer the interested reader to the literature~\cite{emn,yung}.
\pr
To this end, we remind the reader that
the action of $SL(2,R)/U(1)$ coset Wess-Zumino
model \cite{witt}
describing a Euclidean black hole can be written
in the form~\cite{witt}
\be
S=\frac{k}{4\pi} \int d^2z \frac{1}{1+|w|^2}\partial _\mu {\bar w}
\partial ^\mu w + \dots
\label{threev}
\ee
where the conventional radial
and angular coordinates $(r,\theta)$ are given
by $w=sinh r e^{-i\theta}$ and the target
space $(r,\theta)$ line element is
\be
ds^2=\frac{dwd{\overline w}}{1 + w{\overline w}}=dr^2+tanh^2rd\theta^2
\label{fourv}
\ee
The corresponding exactly-marginal deformation,
which turns on matter backgrounds
in this geometry is
constructed by $W_\infty$ symmetry considerations, and
is given by \cite{chaudh}
\be
L_0^1{\overline L}_0^1 \propto
{\cal F}^{c-c}_{\frac{1}{2},0,0} + i(\psi^{++}-\psi^{--}) + \dots
\label{margintax}
\ee
where the $\psi$ denote higher-string-level operators \cite{chaudh},
and
the `tachyon' operator is given by
\be
{\cal F} ^{c-c}_{\frac{1}{2},0,0}(r)
=\frac{1}{coshr}
F(\frac{1}{2},\frac{1}{2} ; 1, tanh^2r )
\label{tachyon}
\ee
with
\bea
&~&F(\frac{1}{2},\frac{1}{2},1;tanh^2r) \simeq
\frac{1}{\Gamma ^2(\frac{1}{2})}\sum_{n=0}^{\infty}
\frac{(\frac{1}{2})_n(\frac{1}{2})_n}{(n !)^2}[2\psi(n+1)-
2\psi(n+\frac{1}{2})+ \nn \\
&+&ln(1 + |w|^2)]
(\sqrt{1 + |w|^2}~)^{-n}
\label{wseven}
\eea
There is an additional marginal deformation, dictated by the
$SL(2,R)$ symmetry structure \cite{chaudh}, which consists
of topological string modes only.
At large
$k$, this  operator  rescales the black hole metric, as
can be seen
from
its contribution to the action of the deformed
Wess-Zumino
$\sigma$-model after the gauge field integration \cite{chaudh},
\bea
gL_0^2 {\overline L_0}^2 \ni \int d^2z \{\partial r {\overline
\partial} r (1-2g csch^2 r -2g sech^2 r) + \nn \\
\partial \theta {\overline \partial}\theta
(sinh^2 r + 2g - \frac{(sinh^2 r + 2g)^2}{cosh^2 r + 2g})\}
\label{chlyk}
\eea
Changing variables $cosh^2r + 2g \rightarrow cosh^2r $ in (\ref{chlyk})
one finds that to $O(g)$ the target space metric is rescaled by an
overall constant.
\pr
Notice that (\ref{tachyon}) plays the r\^ole of the
cosmological constant operator in flat space-times.
In our two-dimensional black hole case,
the analogous  exactly marginal deformation
is not simply the cosmological constant, but
the operator (\ref{margintax}). The latter
consists of  an infinite set
of discrete topological
$W_\infty$ states, which, as argued in ref. \cite{emn},
play an important r\^ole in preserving quantum
coherence of the full string theory, due to
information carried by them during a black hole
evaporation/decay process~\cite{emndec}.
Such states are physical, in that they affect the
renormalization group structure of the theory, and in view
of the interpetation of the RG scale as time in target space,
they also affect the temporal evolution of our system.
\pr
To discuss, at least qualitatively,
their effects in our theory we have argued in ref. \cite{emn}
that one has to consider
the effects of non-perturbative world-sheet
configurations.
As dicsussd in refs. \cite{yung,emn},
the $SL(2,R)/U(1)$  Wess-Zumino coset model
describing a Euclidean black hole also has instantons
given by the holomorphic function
\be
 w(z)=\frac{\rho}{z-z_0}
\label{ninev}
\ee
with topological charge
\be
 Q=\frac{1}{\pi}\int d^2z \frac{1}{1+|w|^2}[{\overline \partial}
{\overline w}\partial      w  - h.c. ]
 =-2 ln(a) + ~const
\label{tenv}
\ee
where $a$ is an ultraviolet cut-off.
The instanton action on the world-sheet  also depends
logarithmically on the ultraviolet cut-off. As in the case of
the more familiar vortex configuration in the Kosterlitz-Thouless
model, this logarithmic divergence does not prevent
the instanton from having important dynamical effects.
In the Bosonic $\sigma$-model, sufficient for our purposes here,
the instanton-anti-instanton vertices take the form \cite{yung}
\be
V_{I{\overline I}}\propto -\frac{d}{2\pi}
\int d^2z \frac{d^2\rho}{|\rho|^4}
e^{-S_0} (e^{(\frac{k[\rho\partial {\overline w} + h.c. + \dots ]}
{f(|w|)}}+ e^{(\frac{k[\rho \partial w + h.c. + \dots]}{f(|w|)}})
\label{elevenv}
\ee
introducing a new term into the effective
action. Making a derivative expansion
of the instanton vertex and taking the large-$k$
limit, i.e. restricting our attention
to instanton sizes $\rho \simeq a$, this new term
has the same form as the kinetic term in (\ref{threev}),
and hence corresponds to a renormalization of
the effective level parameter in the
large $k$ limit:
\be
 k \rightarrow k - 2\pi k^2 d'
\qquad : \qquad
 d' \equiv d\int \frac{d|\rho|}{|\rho|^3}
\frac{a^{2}}{[(\rho/a)^2 + 1]^{\frac{k}{2}}}
\label{twelvev}
\ee
If other perturbations are ignored,
the instantons are irrelevant deformations
and conformal invariance is maintained.
However, in the presence of ``tachyon'' deformations,
$T_0 \int d^2z {\cal F}_{-\frac{1}{2}, 0,0}^{c,c}$
in the $SL(2,R)$ notation of ref. \cite{chaudh},
there are extra logarithmic infinities
in the shift (\ref{twelvev}), that are visible in the dilute
gas and weak-``tachyon''-field approximations.
In this case, there
is a contribution to the effective action of the form
\be
T_0\int d^2z d^2z'<{\cal F}_{-\frac{1}{2}, 0, 0}^{c,c}
(z,{\bar z}) V_{I{\overline I}} (z',{\bar z}')>
\label{tachdeform}
\ee
Using the
explicit form of the ``tachyon''
vertex ${\cal F}$ (\ref{tachyon},\ref{wseven})
given by $SL(2,R)$ symmetry
\cite{chaudh}, it is straightforward
to isolate a logarithmically-infinite contribution
to the kinetic term in (\ref{threev}), associated
with infrared infinities on the world-sheet
expressible in terms of the world-sheet area $\Omega /a^2$
\cite{emn},
\bea
        gT_0 \int d^2z' \int
\frac{d\rho}{\rho} (\frac{a^2}{a^2 + \rho^2})^{\frac{k}{2}}
\int d^2 z \frac{1}{|z-z'|^2}
\frac{1}{1 + |w|^2}
\partial _{z'} w(z')
\partial _{\bar z'} {\overline w}(z') + \dots \nn \\
\ni gT_0 ln \frac{\Omega}{a^2} \int d^2z'
\frac{1}{1 + |w|^2}
\partial _{z'} w(z')
\partial _{\bar z'} {\overline w}(z')
\label{analyticexp}
\eea
Such covariant-scale-dependent contributions
can be attributed to Liouville field dynamics, through
the ``fixed-area constraint'' in the Liouville path
integral \cite{distler}. The
zero-mode part
can be absorbed in a
scale-dependent shift of $k$\cite{emn},
which for large $k >>1 $ may be assumed to exponentiate:
\be
k_R\propto (\frac{\Omega}{a^2})^{(const). \beta ^I T_0 }
\label{thirteenv}
\ee
where $\beta ^I$ is the instanton $\beta$-function \cite{yung}.
In ref. \cite{emn} we gave general arguments
and verified to lowest order that instantons represent
higher-string-level (global) mode effects, enabling us to identify
$\beta^I =-\beta ^T $, where
$\beta^T$ is
the renormalization-group $\beta$-function of a
matter deformation of the black hole\footnote{Notice that this
implies that the matter $\beta$-function has to be computed
in a non-perturbative way, which is consistent with the
exact conformal field theory analysis of ref. \cite{chaudh}.}.
Notice that in (\ref{thirteenv}) both the
infrared and the ultraviolet cut-off scales enter.
We do not distinguish between
infrared and ultraviolet cut-offs in our framework. The physical
scale of the system, which varies along a
renormalization group trajectory, is the dimensionless
ratio of the two, which is identified with the Liouville
field.
\pr
The change in $k$ and the associated change in the
central charge $c=\frac{3k}{k-2}-1$
and the black-hole mass
$M_{bh} \propto (k-2)^{-\frac{1}{2}}$
do not conflict with any general theorems.
An analogous instanton renormalization
of $\theta$ (c.f. $k$) has been demonstrated \cite{pruisk}
in related $\sigma$-models
that describe the Integer Quantum Hall Effect (IQHE), discussed further
in ref. \cite{emn}.
\pr
So far we have dealt with target-space
Euclidean black holes. Although
one could appeal to analytic contiuation
arguments {\it a posteriori}, however it
would be useful to have an understanding
of the Minkowski case, which in the string
framework
of ref. \cite{witt} is obtained by
analyzing the coset Wess-Zumino model
over $SL(2,R)/O(1,1)$.
Instanton renormalization of $k$ can also be seen
in this model.
The $\sigma$-model action
of such a theory contains \cite{yung},
in addition to the action (\ref{threev}),
a total-derivative $\theta$-term
which
can be thought of as a deformation of the
black hole by an ``antisymmetric tensor''
background, which in two dimensions
is a discrete mode as a result of the abelian gauge symmetry.
Its Euclideanized version
has also instanton solutions of the form (\ref{ninev}),
but with {\it finite} action,
which
induce
``Liouville''-time-dependent shifts
to $k$, prior to matter couplings.
This situation is similar to what happens
to the {\it topological} $N=2$
theory, discussed in the text.
The conclusions about the
existence of extra divergences
in correlation functions of discrete
`tachyon' operators persist in the Minkowski
picture, thereby justifying the
absorption of such divergencies by
local renormalization scales (Liouville fields)
on the world-sheet.
\pr
In our case, as we have seen, the instantons
reflect
a shift of the central charge between the matter
and background sectors of a combined matter $+$ black hole theory,
in which the total
central charge is unchanged.
Qualtitatively, their effects have been argued~\cite{emn}
to
correspond to
a combination of world-sheet deformation operators in
the Wess-Zumino model~\cite{witt}, pertaining to global modes
at higher-string levels:
the exactly marginal
operator $L_0^2 {\overline L}_0^2 $ and the rest of
the moduli representing the higher-string-level
$W$-hair~\cite{chaudh,emn},
and the irrelevant part
of the exactly-marginal deformation $L_0^1 {\overline L}_0^1$,
which involves an infinite sum
of massive (global) string operators \cite{chaudh}.
The fact that the $L_0^2{\overline L}_0^2 $ operator rescales
the target-space metric by an overall constant, implies
that
such perturbations have
the same effect as the instanton.
Thus the
instanton represents the effects of global higher-level
string
modes that are related to each other and to massless excitations
by a $W$ symmetry.
Matrix elements of the full exactly
marginal light matter $+$ instanton operator have
no dependence on the ultraviolet cut-off $a$, but the separate
matter and instanton parts do depend on $a$, as we have
seen above.
\pr
Since instantons rescale the target-space metric
and the black hole mass, they may also be used
to represent black hole decay.
This is higher-genus effect in string theory \cite{emndec},
so one should expect that instantons could reflect the contributions
of higher genera. This expectation is indeed supported by
an explicit computation of instanton effects in a dilute-gas
approximation in the presence of dilatons.
This point has been briefly discussed in the text, and
for more details we refer the reader to the
literature~\cite{emn,emndec}.
\pr
\newpage
\pr
\nk {\Large {\bf Appendix B} }
\pr
\nk {\bf  Variational Approach to Soliton Quantization
via Squeezed Coherent States }
\pr
It is the purpose of this
appendix to discuss briefly the formalism
leading to the quantization of the solitonic
states discussed in section 2.
\pr
One assumes the existence of a canonical
second quantized formalism for the $(1+1)$-dimensional
scalar field $u(x,t)$, based on creation and
annihilation operators
$a^\dagger _k$,
$a_k$.
One then constructs a squeezed
vacuum state\cite{tdva}
\be
     |\Psi (t) > = N(t) e^{T(t)} |0>
\qquad ; \qquad T(t) = \frac{1}{2} \int \int
dx dy u(x) \Omega (x,y,t) u(y)
\label{sq1}
\ee
where $|0>$ is the ordinary vacuum state
annihilated by $a_k$,
and
$N(t)$ is a normalization factor to be determined.
$\Omega (x,y,t)$ is a complex function,
which can be splitted
in real and imaginary parts
as
\bea
  \Omega (x,y,t) &=& \frac{1}{2}[G_0^{-1}(x,y) - G^{-1}(x,y,t)]
+ 2i\Pi (x,y,t) \nn \\
G_0 (x,y) &=& <0|u(x)u(y)|0>
\label{sq2}
\eea
The squeezed coherent state for this system can be then defined
as\cite{tdva}
\be
 |\Phi (t) > \equiv e^{iS(t)}|\Psi (t)> \qquad ; \qquad
S(t) = \int_{-\infty}^{+\infty} dx[D(x,t)u(x) - C(x,t)\pi (x)]
\label{sq3}
\ee
with $\pi (x)$ the momentum conjugate to $u(x)$, and
$D(x,t)$, $C(x,t)$ real functions.
With respect to this state $\Pi (x,t)$ can
be considered as a momentum canonically conjugate
to $G(x,y,t)$ in the following sense
\be
         <\Phi (t) | -i\frac{\delta}{\delta \Pi (x,y,t)}
|\Phi (t)> = - G(x,y,t)
\label{sq4}
\ee
The quantity $G(x,y,t)$ represents the modified
boson field around the soliton.
\pr
To determine $C$,$D$, and $\Omega$ one applies
the Time-Dependent Variational Approach
(TDVA)~\cite{tdva} according to which
\be
\delta \int _{t_1}^{t_2} dt <\Phi (t) |(
i\partial _t - H) |\Phi (t) >  = 0
\label{sq5}
\ee
where $H$ is the canonical Hamiltonian of the system.
This leads to a canonical set of  (quantum) Hamilton
equations
\bea
{\dot D}(x,t)&=&-\frac{\delta {\cal H}}{\delta
C(x,t)} \qquad  {\dot C}(x,t) = \frac{\delta {\cal H}}
{\delta D (x,t)} \nn \\
{\dot G}(x,y,t) &=& \frac{\delta {\cal H}}{\delta \Pi (x,y,t)}
\qquad {\dot \Pi}(x,y,t) =\frac{\delta {\cal H}}{\delta
G(x,y,t)}
\label{sq6}
\eea
where the quantum energy functional
${\cal H}$ is given by\cite{tdva}
\be
  {\cal H} \equiv <\Phi (t) | H | \Phi (t) >= \int _{-\infty}
^{\infty} dx {\cal E} (x)
\label{sq7}
\ee
with
\bea
{\cal E} (x) = \frac{1}{2} D^2 (x,t) &+& \frac{1}{2} (
\partial _x  C(x,t))^2  + {\cal M}^{(0)} [C(x,t)] + \nn \\
+\frac{1}{8}<x|G^{-1}(t)|y> &+& 2 <x|\Pi (t) G(t) \Pi (t) |y>
+\frac{1}{2}lim_{x \rightarrow y} \nabla _x\nabla _y <x| G(t) |y>
 - \nn \\
-\frac{1}{8}<x|G_0^{-1}|y> &-& \frac{1}{2}lim _{x \rightarrow y}
\nabla _x \nabla _y <x|G_0 (t)|y>
\label{sq8}
\eea
where we use the following operator notation in coordinate
representation
$A (x,y,t) \equiv <x|A(t)|y>$, and
\be
M^{(n)} = e^{\frac{1}{2}(G(x,x,t)-G_0(x,x))\frac{\partial ^2}
{\partial z^2}} U^{(n)}(z) |_{z=C(x,t)}
\qquad ; \qquad U^{(n)} \equiv d^n U/d z^n
\label{sq9}
\ee
Above, $U $ denotes the potential of the original soliton
Hamiltonian, $H$.
Notice that the quantum energy functional is conserved
in time, despite the various time dependences
of the quantum fluctuations. This is a consequence
of the canonical form (\ref{sq6}) of the Hamilton equations.
\pr
Performing the functional derivations in (\ref{sq6})
one can get
\bea
 {\dot D}(x,t) &=& \frac{\partial ^2}{\partial x^2}
C(x,t) - {\cal M}^{(1)}[C(x,t)]   \nn \\
{\dot C}(x,t) &=& D(x,t)
\label{sq10}
\eea
which after elimination of $D(x,t)$, yields
the modified (quantum) soliton equation (\ref{22c}).
We note that the quantities ${\cal M}^{(n)}$ carry
information about the quantum corrections, and in
this sense tha above scheme is more accurate
than the $WKB$ approximation \cite{WKB}.
The whole scheme may be thought of as
a mean-field-approach to quantum corrections to the soliton
solutions.
\pr
In our string
framework, then,
these point-like quantum solitons can be viewed
as a low-energy approximation to some more general
ground state solutions of a
non-critical string theory, formulated in higher genera on the
world sheet to account for the quantum corrections.
We have not worked out in this work the full string-theory
representation of the relevant quantum coherent state.
This will be an interesting topic to be studied in the future,
which will allow for a rigorous
study of the effects of
the global string modes on the collapse of
the quantum-coherent preconscious state.

\newpage
{\Large {\bf Figure Captions}}
\pr

\nk {\Large {\bf Figure 1 }}- Microtubular Arrangement :
(a) the structure of a Microtubule (MT), (b)
cross section of a MT, (c) two neighboring dimers  along
the direction of a MT axis
\pr

\nk {\Large {\bf Figure 2 }}- The two conformations
$\alpha $ and $\beta$ of a MT dimer.
Transition (switching)
between these two conformational states
can be viewed as a quantum-mechanical effect.
Quantum-Gravity entanglement can cause the collapse
of quantum-coherent states of such conformations,
which might arise
in a MT network modelling the preconscious state of
the human brain.

\pr
\nk {\Large {\bf Figure 3 }}- Illustration of the phenomenon
of `dynamical instability' of a MT network : (a) unbounded
`sawtooth' growth  (b) bounded `sawtooth' growth.
Dotted lines show the average over many MT with the
same dynamical parameters.

\pr
\nk {\Large {\bf Figure 4 }}- (a) Contour
of integration in the analytically-continued
(regularized) version of $\Gamma (-s)$ for $ s \in Z^+$.
This is known in the literature as the Saalschutz contour,
and has been used in
conventional quantum field theory to relate dimensional
regularization to the Bogoliubov-Parasiuk-Hepp-Zimmermann
renormalization method,
(b) schematic repesentation
of the evolution of the world-sheet area as the renormalization
group scale moves along the contour of fig. 4(a)

\end{document}